\documentstyle[epsfig,a4p,12pt,rotating]{article}
\begin{document}
%
%
\newcommand{\PPEnum}    {CERN-PPE/97-121}
\newcommand{\PNnum}     {OPAL Physics Note PN-295}
\newcommand{\TNnum}     {OPAL Technical Note TN-xxx}
\newcommand{\Date}      {2 December 1997}
\newcommand{\Author}    {M.~Oreglia, A.~S.~Turcot}
\newcommand{\MailAddr}  {Mark.Oreglia@cern.ch}
\newcommand{\EdBoard}   {E. Duchovni, A.~F\H{u}rtjes, M.~Roney, R. Teuscher}
\newcommand{\DraftVer}  {Version 2.07}
\newcommand{\DraftDate} {\today}
\newcommand{\TimeLimit} {Monday, 1 September 1997, 08h00 Geneva time}

\def\toprule{\noalign{\hrule \medskip}}
\def\midrule{\noalign{\medskip\hrule }}
\def\botrule{\noalign{\medskip\hrule }}
\setlength{\parskip}{\medskipamount}

\newcommand{\TPONE}{$\theta_{\gamma 1}$}
\newcommand{\TPTWO}{$\theta_{\gamma 2}$}
\newcommand{\CTPM}{$\cos\theta_{\mrm miss}$}
\newcommand{\ECV} {$E_{\mrm{excess}}$}
\newcommand{\NPHO} {$\mrm N_{\gamma} > 1$}

\newcommand{\mgg} {$M_{\gamma \gamma}$}
\newcommand{\mdip} {M_{\gamma \gamma}}
\newcommand{\Stogg} {${\mathrm S} \ra \gamma \gamma$}
\newcommand{\epem}{{\mathrm e}^+ {\mathrm e}^-}
\newcommand{\tptm}{{\tau}^+ {\tau}^-}
\newcommand{\mpmm}{{\mu}^+ {\mu}^-}

\newcommand{\qqbar}{{\mathrm q}\bar{\mathrm q}}
\newcommand{\nn}{\nu \bar{\nu}}
\newcommand{\nunu}{\nu \bar{\nu}}
\newcommand{\mumu}{\mu^+ \mu^-}
\newcommand{\ellell}{\ell^+ \ell^-}
\newcommand{\MZ}{M_{\mathrm Z}}
\newcommand{\MH}{M_{\mathrm H}}
\newcommand{\MX} {M_{\mathrm{X}}}
\newcommand{\MY} {M_{\mathrm{Y}}}

\newcommand {\Hboson}        {{\mathrm H}^{0}}
\newcommand {\Hzero}         {${\mathrm H}^{0}$}
\newcommand {\Zboson}        {{\mathrm Z}^{0}}
\newcommand {\Zzero}         {${\mathrm Z}^{0}$}
\newcommand {\Wpm}           {{\mathrm W}^{\pm}}
\newcommand {\allqq}         {\sum_{q \neq t} q \bar{q}}
\newcommand {\mixang}        {\theta _{\mathrm {mix}}}
\newcommand {\thacop}        {\theta _{\mathrm {Acop}}}
\newcommand {\cosjet}        {\cos\thejet}
\newcommand {\costhr}        {\cos\thethr}
\newcommand{\epair}    {\mbox{${\mathrm e}^+{\mathrm e}^-$}}
\newcommand{\mupair}   {\mbox{$\mu^+\mu^-$}}
\newcommand{\taupair}  {\mbox{$\tau^+\tau^-$}}
\newcommand{\qpair}    {\mbox{${\mathrm q}\overline{\mathrm q}$}}
\newcommand{\ff}       {{\mathrm f} \bar{\mathrm f}}
\newcommand{\gaga}     {\gamma\gamma}
\newcommand{\WW}       {{\mathrm W}^+{\mathrm W}^-}
\newcommand{\eeee}     {\mbox{\epair\epair}}
\newcommand{\eemumu}   {\mbox{\epair\mupair}}
\newcommand{\eetautau} {\mbox{\epair\taupair}}
\newcommand{\eeqq}     {\mbox{\epair\qpair}}
\newcommand{\eeff}     {\mbox{$\mathrm e^+e^- f \bar{\mathrm f}$}}
\newcommand{\llnunu}   {\mbox{\lpair\nu\nubar}}
\newcommand{\lnuqq}    {\mbox{\lept\nubar\qpair}}
\newcommand{\zee}      {\mbox{Zee}}
\newcommand{\wenu}     {\mbox{We$\nu$}}

\newcommand{\el}       {\mbox{${\mathrm e}^-$}}
\newcommand{\selem}    {\mbox{$\tilde{\mathrm e}^-$}}
\newcommand{\smum}     {\mbox{$\tilde\mu^-$}}
\newcommand{\staum}    {\mbox{$\tilde\tau^-$}}
\newcommand{\slept}    {\mbox{$\tilde{\ell}^\pm$}}
\newcommand{\sleptm}   {\mbox{$\tilde{\ell}^-$}}
%
%
\newcommand{\zz}        {\mbox{$|z_0|$}}
\newcommand{\dz}        {\mbox{$|d_0|$}}
\newcommand{\sint}      {\mbox{$\sin\theta$}}
\newcommand{\cost}      {\mbox{$\cos\theta$}}
\newcommand{\mcost}     {\mbox{$|\cos\theta|$}}
\newcommand{\dedx}      {\mbox{$dE/dx$}}
\newcommand{\wdedx}     {\mbox{$W_{dE/dx}$}}
\newcommand{\xe}        {\mbox{$x_{\rm E}$}}
\newcommand{\xgam}      {x_{\gamma}}
\newcommand{\Mrec}      {M_{\mrm{recoil}}}
\newcommand{\Lgg}       {\mbox{${\cal L}(\gaga)$}}
\newcommand{\ggjj}{\mbox{$\gaga\;{\rm jet-jet}$ }}
\newcommand{\ggqq}{\mbox{$\gaga {\rm q\bar{q}}$ }}
\newcommand{\ggll}{\mbox{$\gaga\ell^{+}\ell^{-}$}}
\newcommand{\ggnn}{\mbox{$\gaga\nu\bar{\nu}$}}
\newcommand{\eeeell}{\mbox{\epair$\rightarrow$\epair\lpair}}
\newcommand{\eell}{\mbox{\epair\lpair}}
\newcommand{\llgam}{\mbox{$\ell\ell(\gamma)$}}
\newcommand{\nunugam}{\mbox{$\nu\bar{\nu}\gaga$}}
\newcommand{\acope}{\mbox{$\Delta\phi_{\mathrm{EE}}$}}
\newcommand{\nee}{\mbox{N$_{\mathrm{EE}}$}}
\newcommand{\eesum}{\mbox{$\Sigma_{\mathrm{EE}}$}}
\newcommand{\acoph}{\mbox{$\Delta\phi_{\mathrm{HCAL}}$}}
\newcommand {\mm}         {\mu^+ \mu^-}
\newcommand {\emu}        {\mathrm{e}^{\pm} \mu^{\mp}}
\newcommand {\et}         {\mathrm{e}^{\pm} \tau^{\mp}}
\newcommand {\mt}         {\mu^{\pm} \tau^{\mp}}
\newcommand {\lemu}       {\ell=\mathrm{e},\mu}
\newcommand{\Zz}{\mbox{${\mathrm{Z}^0}$}}
%
\newcommand{\Ecm}         {\mbox{$E_{\mathrm{cm}}$}}
\newcommand{\Ebeam}       {E_{\mathrm{beam}}} 
\newcommand{\ipb}         {\mbox{pb$^{-1}$}}

\newcommand{\gsim}{\;\raisebox{-0.9ex}
           {$\textstyle\stackrel{\textstyle >}{\sim}$}\;}
\newcommand{\lsim}{\;\raisebox{-0.9ex}{$\textstyle\stackrel{\textstyle<}
           {\sim}$}\;}

\newcommand{\degree}    {^\circ}
%
\newcommand{\roots}     {\sqrt{s}}
%
%
\newcommand{\thmiss}    { \theta_{miss} }
\newcommand{\cosmiss}   {| \cos \thmiss |}
%
%
\newcommand{\Evis}      {\mbox{$E_{\mathrm{vis}}$}}
\newcommand{\Rvis}      {\mbox{$R_{\mathrm{vis}}$}}
\newcommand{\Mvis}      {\mbox{$M_{\mathrm{vis}}$}}
\newcommand{\Rbal}      {\mbox{$R_{\mathrm{bal}}$}}
%
%
%
\newcommand{\PhysLett}  {Phys.~Lett.}
\newcommand{\PRL}       {Phys.~Rev.\ Lett.}
\newcommand{\PhysRep}   {Phys.~Rep.}
\newcommand{\PhysRev}   {Phys.~Rev.}
\newcommand{\NPhys}     {Nucl.~Phys.}
\def\NIM                {\mbox{Nucl. Instrum. Meth.}}
\newcommand{\NIMA}[1]   {\NIM\ {\bf A{#1}}}
\newcommand{\IEEENS}    {IEEE Trans.\ Nucl.~Sci.}
\newcommand{\ZPhysC}[1]    {Z. Phys. {\bf C#1}}
\newcommand{\PhysLettB}[1] {Phys. Lett. {\bf B#1}}
\newcommand{\CPC}[1]      {Comp.\ Phys.\ Comm.\ {\bf #1}}
\def\etal{\mbox{{\it et al.}}}
%
%
\newcommand{\OPALColl}  {OPAL Collab.}
\newcommand{\JADEColl}  {JADE Collab.}
%
\newcommand{\onecol}[2] {\multicolumn{1}{#1}{#2}}
\newcommand{\colcen}[1] {\multicolumn{1}{|c|}{#1}}
\newcommand{\ra}        {\rightarrow}   
\newcommand{\ov}        {\overline}   
\def\mrm       {\mathrm}
\newcommand {\downto}
        {\mbox{ \begin{picture}(14,10)
                   \put(0,10){\line(0,-1){5.0}}
                   \put(2,5){\oval(4,4)[bl]}
                   \put(1,0){\makebox(0,0)[bl]{$\rightarrow$}}
                \end{picture} }}


\begin{titlepage}
\begin{center}{\large   EUROPEAN LABORATORY FOR PARTICLE PHYSICS
}\end{center}\bigskip
\begin{flushright}
       CERN-PPE/97-121   \\ \Date
\end{flushright}
\bigskip\bigskip\bigskip\bigskip\bigskip
%
%
\begin{center}{\huge\bf        Search for a Massive Di-photon \\
                               Resonance at $\sqrt{s} = 91-172$ GeV
}\end{center}\bigskip\bigskip
\begin{center}{\LARGE The OPAL Collaboration
}\end{center}\bigskip\bigskip
%
%
\bigskip\begin{center}{\large  Abstract}\end{center}
A search for the resonant production of high mass photon pairs
associated with a leptonic or hadronic system has been performed using 
a total data sample of 25.7~\ipb\ taken at centre-of-mass energies
between 130~GeV and 172~GeV with the OPAL detector at LEP.
The observed number of events
is consistent with the expected number from Standard Model processes.
The observed candidates are combined with search results from 
$\roots \approx \MZ$
to place limits on $B$($\Hboson \ra \gaga$) within the Standard Model
for Higgs boson masses up to 77 GeV,
and on the production cross section of any scalar resonance decaying
into di-photons.
Upper limits on $B$($\Hboson \ra \gaga)\times\sigma(\epem\ra\Hboson\Zboson)$
of 290 -- 830~fb are obtained over $40 < \MH < 160$~GeV.
Higgs scalars which couple only to gauge bosons at Standard Model strength
are ruled out up to a mass of 76.5~GeV at the 95\% confidence level.

\bigskip\bigskip\bigskip\bigskip
\bigskip\bigskip
\begin{center}{\large
(Submitted to Zeitschrift f\"ur Physik C)
}\end{center}
\end{titlepage}
\begin{center}{\Large        The OPAL Collaboration
}\end{center}\bigskip
\begin{center}{
K.\thinspace Ackerstaff$^{  8}$,
G.\thinspace Alexander$^{ 23}$,
J.\thinspace Allison$^{ 16}$,
N.\thinspace Altekamp$^{  5}$,
K.J.\thinspace Anderson$^{  9}$,
S.\thinspace Anderson$^{ 12}$,
S.\thinspace Arcelli$^{  2}$,
S.\thinspace Asai$^{ 24}$,
D.\thinspace Axen$^{ 29}$,
G.\thinspace Azuelos$^{ 18,  a}$
% ...
}\end{center}\bigskip
\bigskip
$^{  1}$School of Physics and Space Research, University of Birmingham,
Birmingham B15 2TT, UK
\newline
$^{  2}$Dipartimento di Fisica dell' Universit\`a di Bologna and INFN,
I-40126 Bologna, Italy
\newline
% ...
$^{  5}$Cavendish Laboratory, Cambridge CB3 0HE, UK
\newline
% ...
$^{  8}$CERN, European Organisation for Particle Physics,
CH-1211 Geneva 23, Switzerland
\newline
$^{  9}$Enrico Fermi Institute and Department of Physics,
University of Chicago, Chicago IL 60637, USA
\newline
% ...
$^{ 12}$Indiana University, Department of Physics,
Swain Hall West 117, Bloomington IN 47405, USA
\newline
% ...
$^{ 16}$Department of Physics, Schuster Laboratory, The University,
Manchester M13 9PL, UK
\newline
% ...
$^{ 18}$Laboratoire de Physique Nucl\'eaire, Universit\'e de Montr\'eal,
Montr\'eal, Quebec H3C 3J7, Canada
\newline
% ...
$^{ 23}$Department of Physics and Astronomy, Tel Aviv University,
Tel Aviv 69978, Israel
\newline
$^{ 24}$International Centre for Elementary Particle Physics and
Department of Physics, University of Tokyo, Tokyo 113, and
Kobe University, Kobe 657, Japan
\newline
% ...
$^{ 29}$University of British Columbia, Department of Physics,
Vancouver BC V6T 1Z1, Canada
\newline
% ...
$^{  a}$ and at TRIUMF, Vancouver, Canada V6T 2A3
\newline
% ...
$^{  d}$ and Department of Experimental Physics, Lajos Kossuth
University, Debrecen, Hungary
\newline
$^{  e}$ and Department of Physics, New York University, NY 1003, USA
\newline
\newpage
\section{Introduction}
\label{sec:intro}

This paper describes a search for a massive di-photon resonance
produced in $\epem$ collisions from $\roots$ = 91 to 172 GeV.
The search presented here is based on a total of 
173~\ipb\ of data taken at $\roots \approx 91$ GeV (``LEP1''),
5.4~\ipb\ taken at $\roots = 130-140$ GeV (``LEP1.5'', also referred to as 133 GeV,
which is the luminosity-weighted energy average), and
20.3~\ipb\ taken at $\roots = 161-172$ GeV (``LEP2'')\footnote{
More precisely, the LEP1.5 and LEP2 datasets consist of
2.73~\ipb\ at 130.3~GeV, 2.64~\ipb\ at 136.2~GeV, 0.05~\ipb\ at 140.1~GeV,
10.0~\ipb\ at 161.3 GeV, 1.0~\ipb\ at 170.3 GeV, and 9.3~\ipb\ at 172.3~GeV.}.

For a hypothetical di-photon resonance produced with \mgg$ > 20$ GeV,
the signature is rather distinct from backgrounds because the photons are so energetic.
At centre-of-mass energies above the \Zzero\,
the most important background arises from initial state radiation leading to
doubly radiative returns to the \Zzero\ 
($\epem \ra \Zboson (\gamma \gamma)_{\rm ISR}$). 
The \ggqq, \ggll, and \ggnn\ final states are
a potentially rich hunting ground for non-Standard Model processes. 
In the case of the Standard Model Higgs boson,
$\Hboson \ra \gaga$ proceeds by means of a vertex loop
and is too small for observation at existing
accelerators even for a kinematically accessible Higgs boson~\cite{HBR}.
An 80 GeV Higgs boson, for example, has an expected di-photon
branching ratio of $1.0\times10^{-3}$. 
However, for anomalous Higgs couplings, the production cross section 
and/or the branching ratio could be large~\cite{Hagiwara}. 
Of particular interest are so-called Type I Two-Higgs doublet models 
where one of the doublets couples only to the $SU(2)\times U(1)$ gauge 
bosons giving rise to a ``bosophilic" scalar~\cite{typeI}. 
Other particles indicative of physics outside the Standard Model
might have distinctive signatures in the di-photon decay mode.

There are existing limits 
on the production of a di-photon resonance which couples to the \Zzero\
from data taken at 
$\roots \approx 91$ GeV from 1991--1994~
\cite{OPAL_ggjj,OPAL_ggll,other_gg},
and measurements of $\gaga \nunu$ at LEP1.5 
and at LEP2 have been published~\cite{ggnunu}.
This paper describes the search for a di-photon resonance produced via
the process $\epem \ra \mrm X Y$, $\mrm X \ra \gaga, Y \ra \ff $
where $\ff$ may be quarks, charged leptons, or a neutrino pair. 
For the hadronic final state, no requirement is imposed
on the mass recoiling from the di-photon system, hence the search is
sensitive to any production of the sort $\epem \ra \mrm X Y$, 
$\mrm X \ra \gaga, Y \ra hadrons$. 
In order to assess measurements made at different values of $\roots$,
the data must be analyzed in the context of a production model, therefore
the Standard Model and 2-doublet type Higgs models are used in this
analysis; in this paper, ``\Hzero'' refers to the lightest neutral scalar
where doublet models are discussed.
Both the LEP1 data and those taken at higher energies contribute significantly
to the searches. The larger dataset at LEP1 energies allows for better
limits on the cross section for particle masses below approximately 80 GeV,
but the final state \Zzero\ is off mass-shell. The higher energy datasets
have lower integrated luminosity, but benefit from the presence of an on-shell
\Zzero.
%
%
\section{The OPAL Detector}

The OPAL detector is described in detail elsewhere \cite{detector}; 
therefore, only the sub-detectors important for this analysis will be
described.  
The electromagnetic calorimeter (EC) consists of lead-glass blocks
of two geometries.

The ``barrel" section of the electromagnetic calorimeter covers
the polar region $|\cos\theta| < 0.82$, where
the polar angle $\theta$ was defined with respect to the incident
electron beam direction.
In the barrel region, the lead glass calorimeter blocks are
$24.6$ radiation lengths thick, with each block subtending
an angular region of approximately $40\times 40$ mrad$^2$.
The ``endcap" sections extends the coverage of the polar region to include
$ 0.81 < |\cos\theta| < 0.98$.
In the endcap region, the lead glass calorimeter blocks are
approximately $22$ radiation lengths thick, with
approximately the same angular segmentation as the barrel.

Charged track (CT) reconstruction was
achieved using a system of cylindrical tracking detectors contained
in a uniform 0.435 T magnetic field.
The tracking device central to this analysis was the jet chamber.
For the polar angle range
$| \cos\theta |<0.92 $, charged tracks are reconstructed with nearly
100\% efficiency.

For this analysis, the central  
jet chamber, endcap and barrel electromagnetic calorimeters were required 
to be fully operational.  The most important detector properties for this
analysis were the photon angular and energy resolutions,
which yielded a di-photon invariant mass resolution (RMS) approximately equal
to $\sigma_{\mdip} = 0.42$~GeV  + $0.02\mdip$, on average,
for scalar production in the energy range considered in this paper.

The quality of reconstruction of
electromagnetic clusters and the accuracy on the modelling of backgrounds
varied in several ranges of the polar angle.
The polar angle range $0.82 > |\cos\theta| > 0.81$
is the region of overlap between the barrel and endcap electromagnetic calorimeters;
electromagnetic clusters are not as well measured in this region.
For $0.8 > |\cos\theta| > 0.7$, material from the jet chamber pressure vessel
somewhat degrades photon and electron energy measurement.
%
Inert material in the polar angle range $|\cos\theta| < 0.9$ is well modelled
in the Monte Carlo simulation of the OPAL detector;
therefore, the polar angle of candidate photons is required to be
in this range.

\section{Simulation of Signals and Backgrounds}

The background sources were modelled by a number of 
different Monte Carlo simulation programs. The Standard Model backgrounds from
$\epem \ra (\gamma/{\rm Z)^{\ast} \ra {\rm q\bar{q}}} $ were simulated
using the PYTHIA \cite{PYTHIA} package with the set of hadronization 
parameters described in reference \cite{jtparams}. Hadronic 
4-fermion processes were modelled using the grc4f \cite{grc4f} and 
EXCALIBUR \cite{excalibur} event 
generators~\footnote{The EXCALIBUR and grc4f results were compared within the context
of this analysis and found to agree within statistical uncertainty.}.
%
%
The process $\epem \ra \gaga(\gamma)$ was simulated using the
RADCOR generator \cite{RADCOR}. The programs BHWIDE
\cite{BHWIDE} and TEEGG \cite{TEEGG} were utilized
to model the background from Bhabha scattering. 
The processes $\epem \ra \ellell$ 
with $\ell \equiv \mu , \tau$ were simulated using
KORALZ \cite{KORALZ}. The KORALZ program was also used to 
generate events of the type $\epem\ra\nu\ov{\nu}\gamma(\gamma)$.
Four-fermion processes of the type 
$\epem \ell^+ \ell^-$, where $\ell \equiv {\rm e},\mu,\tau$, were
modelled using the Vermaseren \cite{VERMASEREN} and grc4f 
generators. The background contributions from the process
$\epem \ra \epem \qqbar$ were simulated using PYTHIA and
HERWIG~\cite{HERWIG}.

For the simulation of potential signals, both the HZHA generator \cite{HZHA} and
the PYTHIA generator were used to 
simulate the process of $\epem \ra \Hboson\Zboson$ followed
by $\Hboson \ra \gaga$ for each \Zzero\ decay channel. 
For the more general production of scalar/scalar and scalar/vector production,
$\epem \ra {\rm XY} \ra \gaga~+~hadrons$, a mass grid was generated. 
For each X or Higgs mass, production samples of 1000 events were generated
from $\MX$,$\MY$ = (40,40) GeV, in 20 GeV steps 
forming an X-Y mass grid,
up to the kinematic limit
for each of the LEP2 centre-of-mass energies.

For simulation of the 1995 LEP1 data used in this analysis, the
JETSET 7.4~\cite{JETSET} and HERWIG 5.8 programs
were used. The JETSET program appears to simulate the production
of photons and neutral particles better than HERWIG, though both
programs underestimate the numbers of low energy isolated photons
and $\pi^{0}$ mesons~\cite{BADPI}. 

Both signal and background events were processed using the full
OPAL detector simulation~\cite{GOPAL}.
The detector simulation describes the data well except for the
low polar angle region mentioned in the previous section.

\section{Event Selection}

The philosophy adopted in this analysis was to introduce the minimum number 
of cuts which allow for a relatively uniform acceptance over the largest 
possible range of masses. 
The search was divided into three topologies.
The first was a search for a system of two photons with large invariant mass
recoiling from a hadronic system.  
The second topology was
a search for di-photons produced in association
with a \Zzero\ decaying to charged leptons.
The third topology was a search for no significant detector activity
other than a di-photon pair. 
Backgrounds in the cases of the charged lepton and missing energy channels
required that the search in these channels be restricted to the case
where the di-photon system recoiled from a \Zzero\ or that the di-photon
energies were less that $\roots$. 
However, the exceptionally clean nature of the di-photon final states permitted
the use of very loose selection criteria to identify the \Zzero\ decay
products.

Radiative events were distinguished by examining
the polar angle distribution of the photons.  Photons arising from 
initial state radiation are close to the beam
direction, whereas photons from processes of interest, 
{\em i.e.}, $\mrm {X} \ra \gaga$, 
would be distributed nearly isotropically. 
The background is serious for photon energies below approximately 10 GeV, 
corresponding to masses below about 20--30~GeV for the centre-of-mass 
energies under consideration.

\subsection{Photon Identification}
\label{s:photid}

Photon candidates were initially selected as ``unassociated''
electromagnetic calorimeter clusters, where no CT track 
was reconstructed within the resolution of the EC cluster.
To make the photon selection more robust, cuts were made on the 
lateral spread and isolation of the electromagnetic clusters. Good clusters were 
required to have lateral sizes consistent with electromagnetic showers. 
The number of blocks in the cluster ($N_{\mrm{blk}}$)
and the number of blocks containing 90\% of the cluster energy 
($N_{90}$)
were required to be less than some maximum values, 
depending on the polar angle of the cluster.
The barrel and endcap regions of the calorimeter described in Section 2
were treated somewhat differently,
and the barrel region was divided into two regions because of differing
amounts of inert material in front of the electromagnetic calorimeter 
in these regions.
Clusters containing channels having excessive readout
noise were eliminated.
The cluster definition cuts were:
\begin{itemize}
\item Barrel region I ($|\cos\theta| < 0.7$): $N_{\mrm{blk}}<15,N_{90}<3$;
\item Barrel region II ($0.7 < |\cos\theta| < 0.81$): $N_{\mrm{blk}}<25,N_{90}<4$;
\item Barrel-Endcap overlap ($0.81 < |\cos\theta| < 0.82$): $N_{\mrm{blk}}<35,N_{90}<5$;
\item Endcap ($0.82 < |\cos\theta| < 0.98$): $N_{\mrm{blk}}<20,N_{90}<5$.
\end{itemize}

\begin{figure}[!htb] 
        \vspace{0.8cm}
        \begin{center}
            \resizebox{\linewidth}{!}{\includegraphics{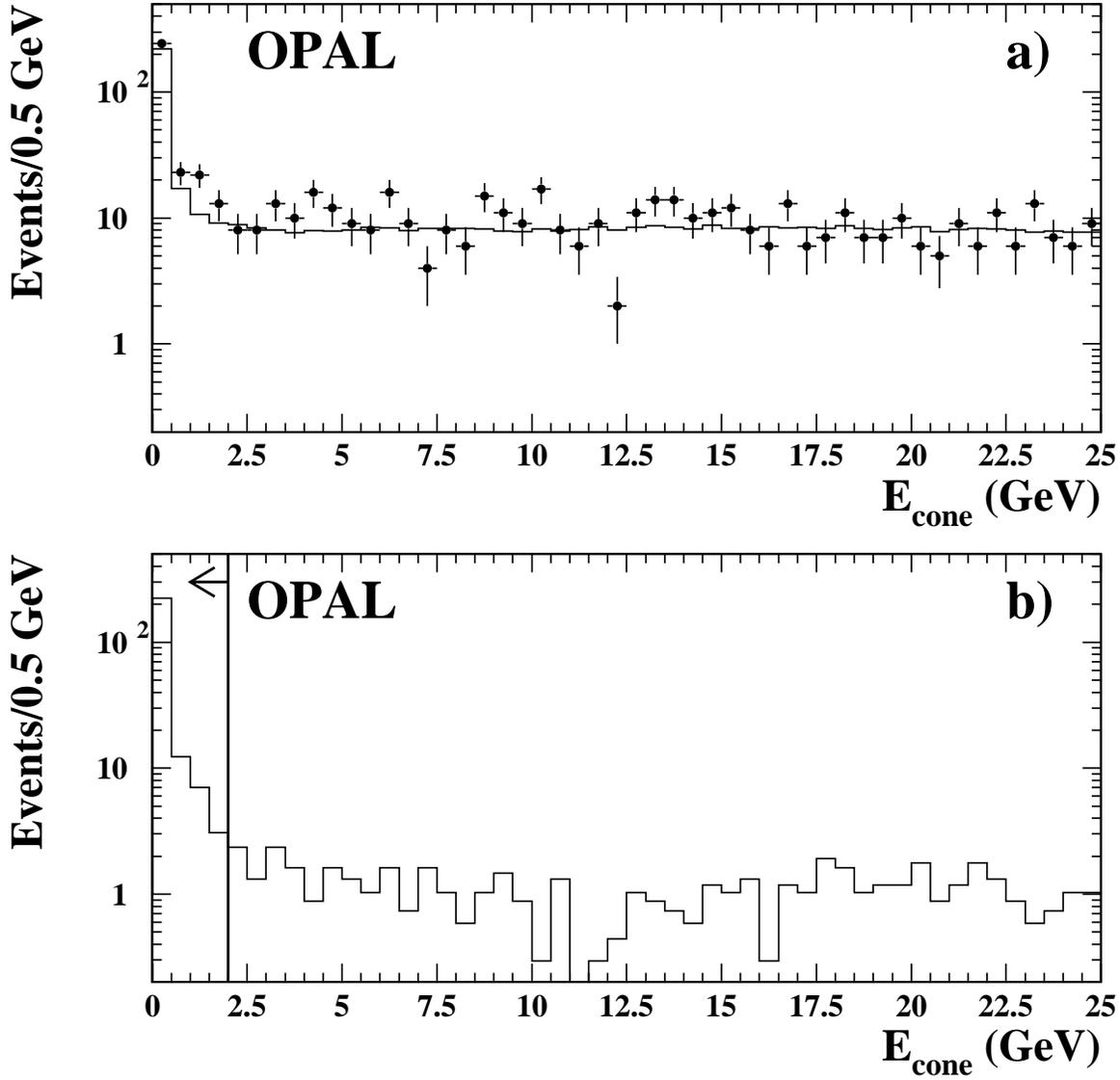} }
        \caption[ECONE]{    
                  Distribution of charged-particle momentum 
                  and unassociated electromagnetic energy sum in $15\degree$ cones about the photon axes
                  (for the hadron channel after multiplicity preselection).
                  (a) 161 GeV data (points) and simulated background (histogram).
                  (b) HZ production with $\MH = 40$ GeV.
                  The position of the cut is shown.
        \label{ECONE} }
        \end{center}
    \end{figure}

Photon candidates were then required to satisfy an isolation
requirement that rejected events where the electromagnetic cluster energy
included particles from the hadronic system.
The energy of additional tracks and clusters
in a $15\degree$ half-angle cone defined by the photon direction had to be less than 
2.0 GeV. 
The distribution of cone energy,
after the multiplicity preselection cuts described in the next section,
is shown in Figure~\ref{ECONE}; the distribution of this variable is also shown for the
simulated background events.
The cone-energy cut reduced the efficiency for signal events by up to 10\%
due to overlap of the photons with particles from the recoil system.
The photon candidates were rejected if there was excessive hadronic
energy behind the electromagnetic cluster; hadron calorimeter energy
within the photon-defining cone had to be less that 20 GeV.
On average, approximately 7\% of the photons converted in material in front of the jet chamber,
producing tracks in the chamber, and were therefore vetoed in this analysis.

\subsection{Hadronic Channel}

\begin{figure}[!htb] 
        \vspace{0.8cm}
        \begin{center}
            \resizebox{\linewidth}{!}{\includegraphics{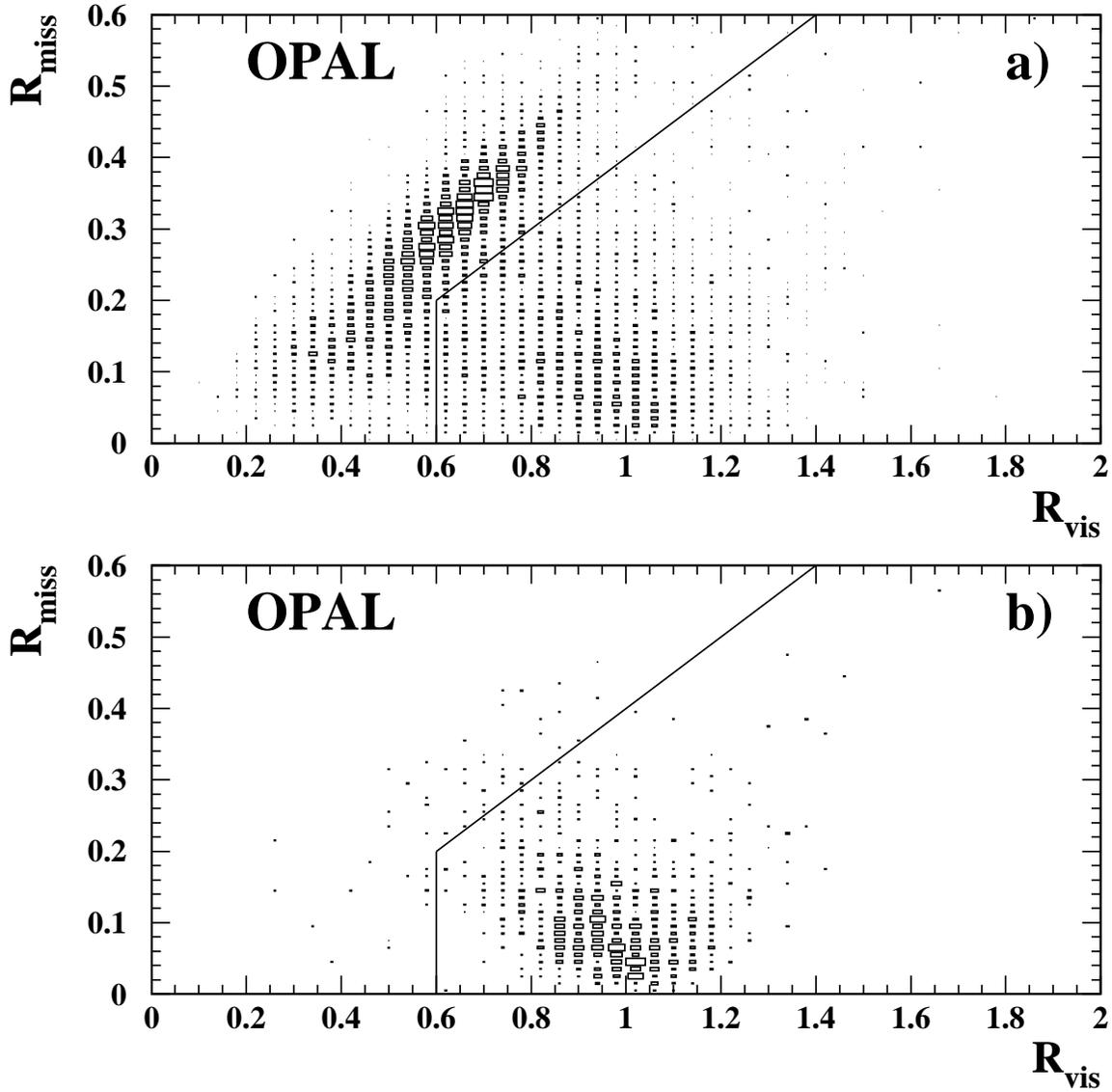} }
        \caption[RVRM]{    
                  Distribution of fractional visible total energy versus
                  fractional missing momentum for 
                  (a) simulation of $\qqbar$ events
                  at $\sqrt{s}$ = 161 GeV, and 
                  (b) simulation of Higgs events
                  with $\MH = 40$ GeV. The cut used for the hadronic channel
                  is shown by the solid line.
        \label{RVRM} }
        \end{center}
    \end{figure}

The hadronic channel consisted of a $\gaga~+~hadrons$ final state. 
Candidates for this topology were initially identified 
by applying a multiplicity preselection
consisting of loose charged track multiplicity and visible energy cuts
which were used in the standard hadronic event selection
described in reference~\cite{hadsel}.
The preselection cuts were applied to the following measured quantities:
\begin{itemize}
\item \Ecm\ $\equiv 2\times \Ebeam$;
\item $\Evis$: sum of CT track energy, unassociated EC, 
               and unassociated hadron calorimeter clusters;
\item $\Rvis \equiv \frac{\mbox{\Evis}}{\Ecm}$;
\item $\vec{p}_{\mrm{vis}}$: vector sum of CT tracks, unassociated EC clusters, 
               and unassociated hadron calorimeter clusters;
\item $R_{\rm miss} \equiv \frac{\mbox{$|p_{\rm{vis}}|$}}{\Ecm}$.
\end{itemize}
The multiplicity preselection cuts required the event to have at least 5 charged tracks
and $\Rvis > 0.1$.
Additional ``precuts" rejected radiative and $\eeff$ events 
using the quantities $\Rvis$ and $R_{\rm miss}$:
\begin{itemize}
\item $\Rvis > 0.6$~and~$R_{\rm miss} < (0.5 \times \Rvis-0.1)$;
\item sum of the visible momentum along the beam direction: 
$|\Sigma~p_{z}^{\mrm{vis}}| < 0.5 \times \Ebeam $;
\item event had to have at least 2 electromagnetic clusters with $E > 0.05\times \Ebeam$.
\end{itemize}
The distributions of $R_{\rm miss}$ and $\Rvis$ for simulated signal
and backgrounds are shown in Figure~\ref{RVRM};
the effects of the cuts on data and background simulations
are shown in Table~\ref{T:qq1}. 

\begin{figure}[!htb] 
        \vspace{0.8cm}
        \begin{center}
            \resizebox{\linewidth}{!}{\includegraphics{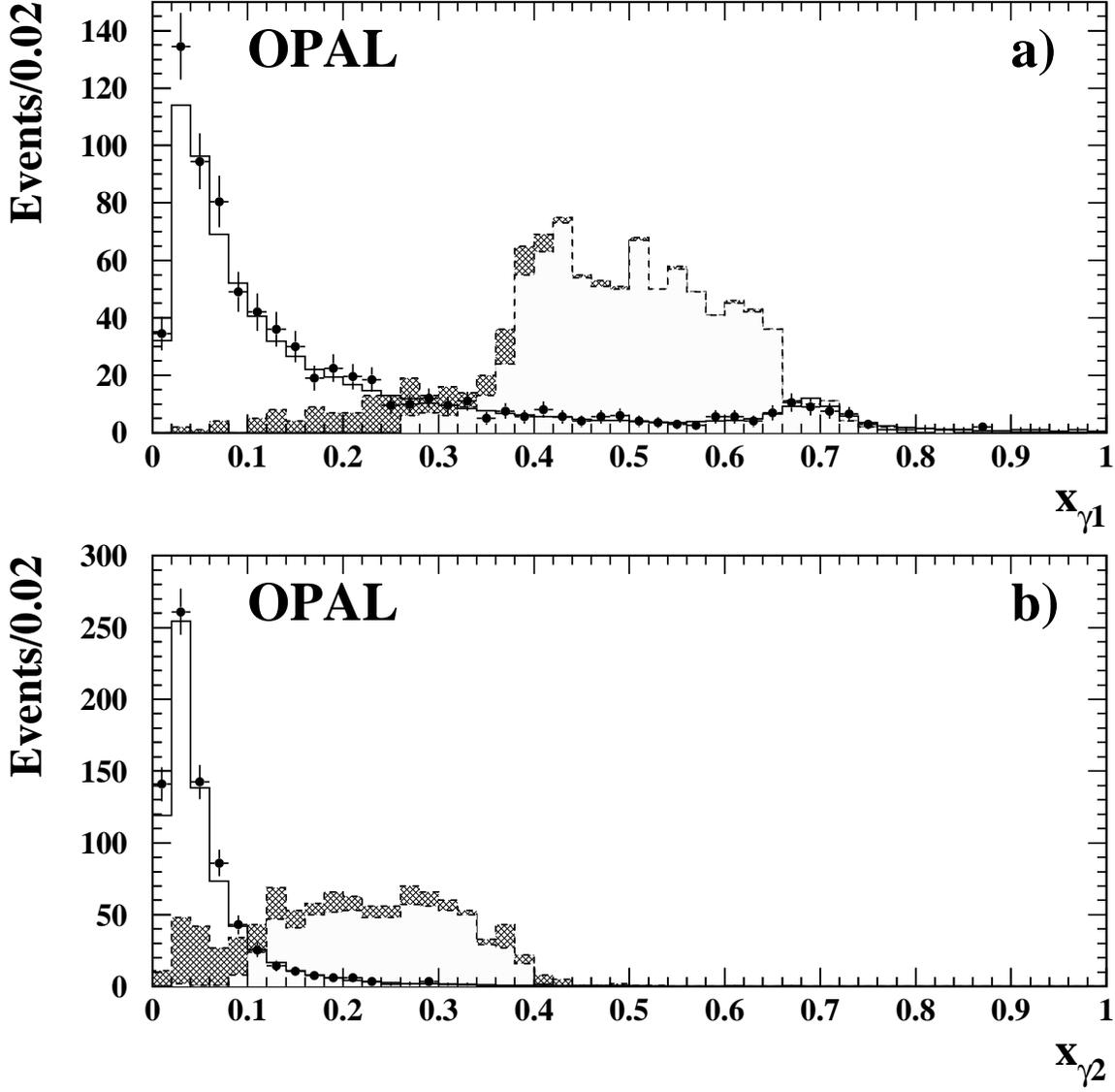} }
        \caption[X1]{    
                  Distribution of $\xgam~\equiv~E_{\gamma}/E_{\mrm beam}$
                  for the most energetic photon (a) and 
                  the second-most energetic photon (b)
                  in the ($\gaga~+~hadrons$) search,
                  after the multiplicity preselection.
                  Data from $\roots$ = 161 GeV
                  are shown as points with error bars; background simulation is
                  indicated by the histogram. The broken histogram shows 
                  \Hzero\Zzero\ production with $\MH = 40$ GeV.
                  The hatched histogram shows simulation cases where the 
                  selected electromagnetic cluster was not due to 
                  the photon from the Higgs boson.
        \label{X1} }
        \end{center}
    \end{figure}

\begin{figure}[!htb]
        \vspace{0.8cm}
        \begin{center}
            \resizebox{\linewidth}{!}{\includegraphics{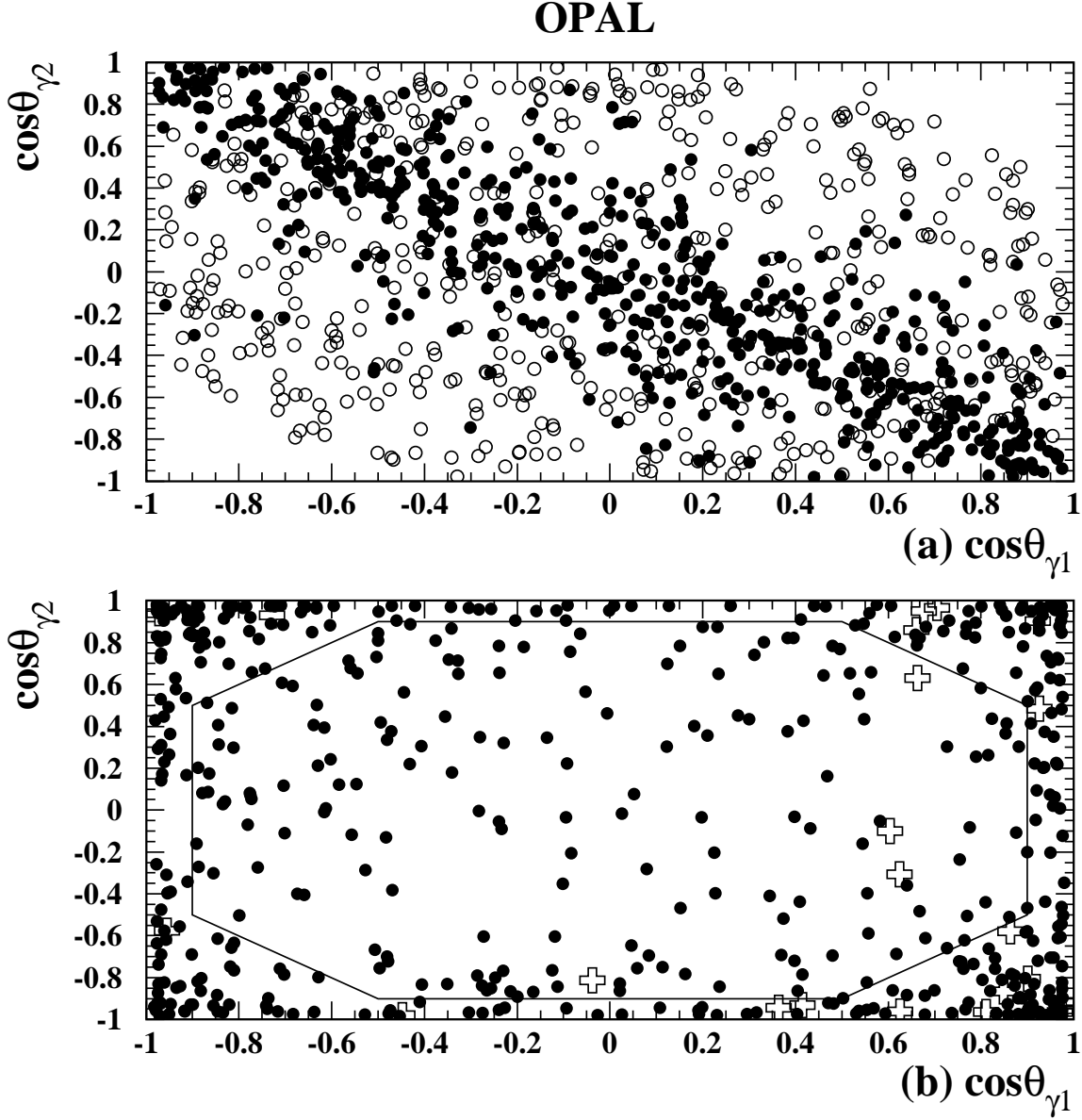} }
        \caption[C1C2S161]{    
                  Distribution of $\cos\theta_{\gamma1}$
                  and $\cos\theta_{\gamma2}$ for simulation
                  events of \Hzero\Zzero\ production at $\roots = 161$ GeV
                  in the hadronic search channel;
                  the precuts have been applied.
                  (a) shows simulated signal for $\MH = 40$ GeV (open circles)
                  and $\MH = 70$ GeV (solid dots).
                  (b) shows simulated $\qqbar$ events
                  and the graphical cut boundary used in the hadronic channel.
                  The data ($\roots$ = 161 and 172 GeV) are shown as open crosses.
        \label{C1C2S161} }
        \end{center}
    \end{figure}

At this point, the background events were almost exclusively from
radiative events, predominantly at low energies and large $|\cos{\theta}|$.
Figure~\ref{X1} shows the distribution for $\xgam$ in data and simulated backgrounds,
as well as for a potential Higgs signal,
where $\xgam $ is defined as $E_{\gamma}/E_{\mrm beam}$,
after applying the multiplicity preselection cuts described in the next section.
In the case of simulated signal, the figure indicates cases in which one of the selected EC
clusters was not from the correct photon. The incidence of such
misidentification falls nearly to zero after more cuts were applied.
As indicated in Table~\ref{T:qq1}, there was a dramatic reduction
of the backgrounds from all sources simply when two energetic photons
were required in the event.  An optimal acceptance for the search
topology was obtained by imposing cuts on the scaled photon energy:
\begin{itemize}
\item Require at least one photon with $x_{\gamma} > 0.10$ , and
\item require at least two photons with $x_{\gamma} > 0.05$ .
\end{itemize}
%
The key difference between the doubly-radiative photons and those
arising from a massive-particle decay is seen in the polar angle distributions
of the photons as shown in Figure~\ref{C1C2S161}.
A cut was therefore imposed to eliminate most of the doubly-radiative events:
\begin{itemize}
\item $|\cos\theta_{\gamma1,2}|<0.9$ and 
$|\cos\theta_{\gamma1}|+|\cos\theta_{\gamma2}|<1.4$.
\end{itemize}
After the cuts on $\theta_{\gamma}$, the agreement between data
and background simulations (Table~\ref{T:qq1}) was good.
Ten events in the LEP1.5 and LEP2 data satisfied all cuts at this point,
which can be compared to the Standard Model
expectation of 8.3~$\pm$~0.5 events (simulation statistical error).
The efficiency for this analysis to accept \Hzero \Zzero\ events 
for $\MH = 40$ and $70$ GeV
is shown in Table~\ref{T:qq1}.
Throughout the mass range of interest, an efficiency greater
than 45\% was maintained.

\subsection{Charged Leptonic Channel}
\label{s:llgg}

The exceptionally clean nature of the $\gaga \ellell $ final state
obviated requiring well-identified leptons. 
As in the hadronic channel, the most serious background for this channel 
was doubly radiative returns to the \Zzero. Bhabha scattering with
initial and/or final state radiation was also a potential background.

Isolated electromagnetic calorimeter clusters and charged tracks satisfying
the selection criteria described in reference \cite{CTSEL} were used to
select charged lepton candidates. To achieve a high efficiency for hadronic 
$\tau$ decays and to ameliorate the possible effects of final state radiation, 
the selected tracks and clusters were combined 
into jets using the Durham recombination scheme \cite{Durham} evaluated 
with $y_{\mrm{cut}} = 0.02$. Candidates were required to have at least two jets
with the possibility of one track defining a jet. The two highest energy 
electromagnetic clusters satisfying the isolation and cluster quality criteria 
of Section \ref{s:photid} were not included. 
No distinction between 
the $\mrm e,\mu$ and $\tau$ channels was made.

%

Leptonic channel candidates were required to satisfy the following
basic selection criteria (referred to as $\ell\ell\gaga$ preselection): 
\begin{itemize}
   \item Low multiplicity preselection requirements~\cite{lowmsel};
   \item precuts particular to the leptonic channel:
 \begin{itemize}
   \item visible energy fraction: $\Rvis>0.2$;
   \item number of EC clusters not associated with tracks:  $2 \leq N_{\rm EC} \leq 10$;
   \item number of good (\cite{CTSEL}) charged tracks: $2 \leq N_{\rm CT} \leq 7$;
   \item momentum fraction along the beam direction:
         $|\Sigma~p_{z}^{\mrm{vis}}|<0.7 \Ebeam$.
\end{itemize}
\end{itemize}
The following additional criteria were then imposed:
\begin{itemize}
\item At least two EC clusters having 
$|\cos{\theta}| < 0.966$
and
         $\xgam >0.1$ satisfying the cluster quality of section \ref{s:photid};
\item at least two jets found (excluding the photon candidates) within the 
           Durham scheme using $y_{\mrm{cut}} = 0.02$.
\end{itemize}

           To further reduce the background from doubly radiative 
           returns, a likelihood selection based on the photon polar 
           angle distributions was utilized. The relative likelihood of
           the di-photon system to be consistent with $\Hboson\ra\gaga$ 
           was defined as:
$$
           {\cal L}(\gaga) = \frac{\displaystyle L(s)}{\displaystyle L(s)+ L(b)} ,
$$
           where $s$ and $b$ referred to signal and background respectively,
           and  
$$
           L(x) = \prod_{i=1,2} P(|\cos\theta_{\gamma_i}|), (x = s,b),
$$
           where $P(|\cos\theta_{\gamma_i}|)$ was the probability of observing 
           photon $i$ at a given $|\cos\theta|$. The reference distributions 
           for the background were taken from $\epem \ra \ff \gamma \gamma$ simulations,
           where ${\rm f} \equiv \mu , \tau, \nu$;
           the electron channel was not used because of the
           t-channel Bhabha process. 
           For the signal distributions,
           $\Hboson\Zboson$ production was assumed with Higgs masses ranging
           from 30 to 80 GeV. The $|\cos\theta|$ distribution exhibited 
           negligible dependence upon the Higgs mass and $\roots$.

Finally, the events had to pass the following two cuts: 
\begin{itemize}
\item Di-photon likelihood: \Lgg\ $>$ 0.4;
\item recoil mass consistent with the \Zzero\ mass:
           $|\Mrec - \MZ| < 20$ GeV, where the recoil mass was computed 
           as that against the di-photon system.
\end{itemize}

\begin{figure}[!p]
        \vspace{0.8cm}
        \begin{center}
            \resizebox{\linewidth}{!}{\includegraphics{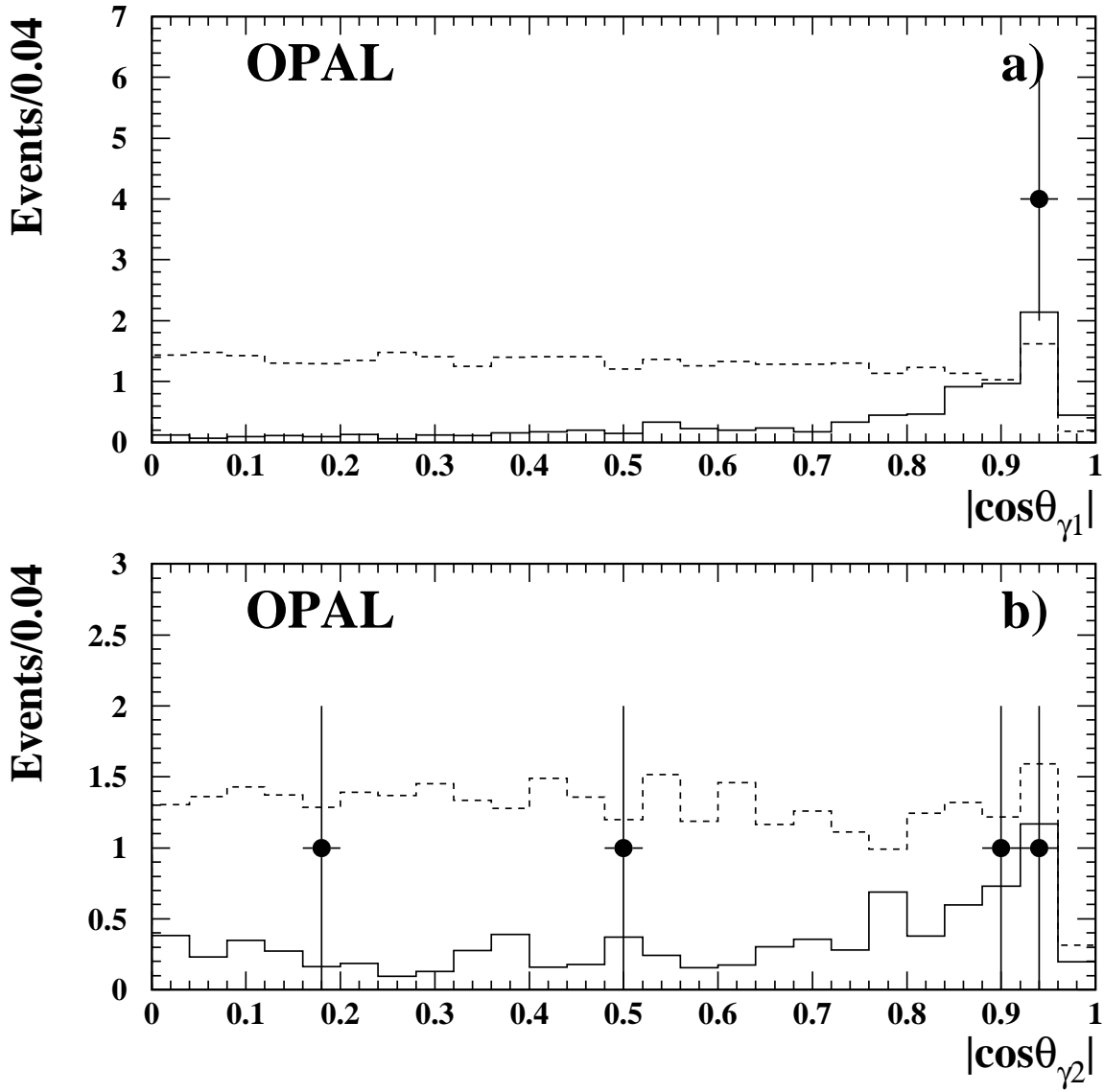} }
        \caption[FLL]{    
                  Distribution of photon polar angles for lepton channel
                  LEP1.5 and LEP2 event candidates, before the likelihood cut.
                  The highest energy photon is shown in a);
                  the lower energy photon is shown in b).
                  Background simulation is indicated by the
                  solid histogram. 
                  The distribution for a 70 GeV
                  Higgs boson is indicated by the broken histogram.
        \label{FLL} }
        \end{center}
    \end{figure}

The cut on the mass recoiling against the di-photon system achieves
a rejection factor of at least 2, as seen in Table~\ref{T:ggll}, for 
a corresponding 5 to 10\% loss of acceptance. 
For events passing the cuts before that on the photon likelihood,
the distribution of photon angles is shown in Figure~\ref{FLL}.
No candidate events were selected at any of the LEP1.5 and LEP2 energies. 
The contribution from Standard Model processes
after the application of all  selection criteria was 
$1.6\pm0.2$, where the error is due to simulation statistics.
The analysis is summarized in Table~\ref{T:ggll}, where
the expected background from leptonic and \eeff\ 4-fermion final states
is compared to the observed number of events. The acceptance for 
$\Hboson\ra\gaga$ ranged from 43--48\% for different Higgs masses. 
                                              
\subsection{Missing Energy Channel}
\label{s:nngg}

The missing energy channel was characterized by a pair of photons recoiling
against a massive, unobserved system. The only Standard Model process
expected to contribute was doubly radiative return to the \Zzero\, followed
by $\Zboson \ra \nu\bar{\nu}$. Potential physics backgrounds included
$\epem \ra \gaga(\gamma)$ and radiative Bhabha scattering with
one or more unobserved electrons.  
Backgrounds due to cosmic rays and beam-wall and beam-gas interactions
were dealt with as described in reference \cite{photsel}. 
Candidates were then required to 
satisfy in addition the following basic selection criteria (referred to as $\nn\gaga$ preselection):
\begin{itemize}
\item 2 electromagnetic clusters with  $\xgam >0.1$, satisfying the cluster 
      quality and isolation criteria described in section \ref{s:photid};
\item $|\Sigma~p_{z}^{\mrm{vis}}| < 0.75 E_{\mrm beam}$;
\item sum of the scaled photon energies: $x_{\gamma1}+x_{\gamma2}~<~1.4$;
\item direction of event missing momentum: $|\cos\theta_{\mrm miss}| < 0.96$;
\item charged track veto: events were required to have no charged track
       candidates consistent with originating from the interaction point and
       having 20 or more jet chamber hits. For tracks at $|\cos\theta|>0.948$,
       the requirement on the number of hits was relaxed to 50\% of the 
       maximum possible hits, up to a minimum of 10 hits. 
       The distance of closest 
       approach to the interaction point in the plane transverse 
       to the beam direction had to be less than 2~cm and the
       distance along the beam axis at this point, $|z_0|$, had to be less than
       50~cm.  To supress backgrounds from beam-gas or beam-wall interactions, 
       events containing one or more tracks with $|z_0|>50~$cm and having 
       at least 20 jet chamber hits were rejected. 
\item excess calorimeter energy (\ECV): the energy observed in the 
      electromagnetic
      calorimeter not associated with the 2 photons was required to
      be less than 3 GeV.
\end{itemize}
At this point the sample was dominated by Bhabha events at small 
polar angles.  The following additional cuts were applied:
\begin{itemize}
\item For each photon polar angle: 
$|\cos{\theta}| < 0.966$;
\item di-photon likelihood: \Lgg$>0.4$. The photon candidates were required to pass the 
      likelihood selection described in Section \ref{s:llgg}.
\end{itemize}

The cut on the sum $x_{\gamma 1}+ x_{\gamma 2}$ 
addresses background from 
$\epem \ra \gaga$ which gives a peak at $x_{\gamma 1}+ x_{\gamma 2}=2$. 
A cut on the energy of the di-photon system was found to be more 
effective than a cut on the acoplanarity angle.
Consequently, this channel is only sensitive to di-photon invariant
masses up to approximately $0.7\roots$.

\begin{figure}[!p]
        \vspace{0.8cm}
        \begin{center}
            \resizebox{\linewidth}{!}{\includegraphics{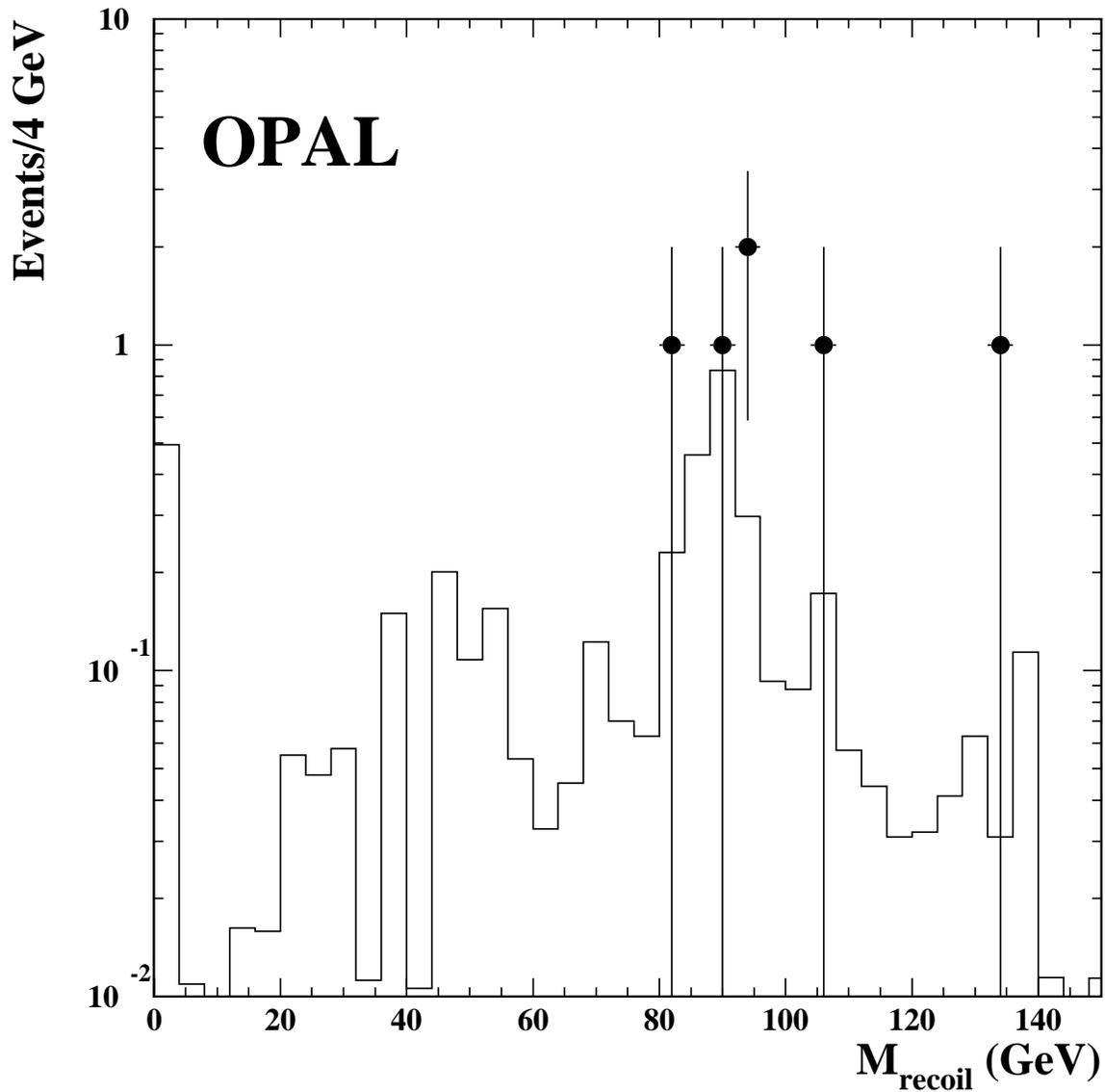} }
        \caption[FNN]{    
                  Recoil mass for missing energy channel LEP1.5 and LEP2 event candidates,
                  before the cut on likelihood.
                  Background simulation is indicated by the solid
                  histogram.
        \label{FNN} }
        \end{center}
    \end{figure}

Three candidates were selected by these cuts in the LEP1.5 and LEP2 data;
this is consistent with the Standard Model 
expectation of $1.8\pm0.2$ events,
where the error is due to simulation statistics.
The level of the simulated background is dominated by the process
$\epem \ra \nunu\gaga$, which has been estimated using
KORALZ~\cite{KORALZ}. 
%
The distribution of recoil mass for these events,
prior to the photon likelihood cut, 
is shown in Figure~\ref{FNN}.
A summary of the effect of the cuts is  given in Table~\ref{T:ggnn}
where the efficiencies for Higgs masses of 40 and 70 GeV are also given.
The acceptance varied from 42 to 64 \% depending on the Higgs mass
and centre-of-mass energy. 

\section{LEP1 Analysis}

Earlier searches 
for the production of a scalar resonance coupling to the \Zzero\
have been performed using
the OPAL detector~\cite{OPAL_ggjj,OPAL_ggll}.
For these analyses at $\roots \approx 91$ GeV, there is
a large background for di-photon invariant masses below approximately
40~GeV.  In the hadron channel, a large component of this background arises from
radiative photons from the initial and final states, and
decays of isolated $\pi^{0}$ and $\eta$ mesons. The
current hadronization simulations JETSET and HERWIG
underestimate the rate of this background. 
Consequently, the LEP1 and LEP2
analyses are compared only for $\mdip > 40$ GeV.

In reference \cite{OPAL_ggll}, $\ellell\gaga$ $(\ell = \rm e, \mu, \tau)$ 
and $\nunu\gaga$  final states were investigated. From a data sample 
consisting of 43~\ipb, corresponding to 1.44 million observable \Zzero\
decays, 2 candidates with \mgg$>40$ GeV were selected in the $\epem\gaga$ 
channel and 2 candidates in the $\mm\gaga$ channel. The background expected 
from the dimuon channel was 1.2 $\pm$ 0.3 events.\footnote{
An evaluation
of the expected background was only available for the muon channel
due to the lack of availability of an event generator for $\epem$
with multiple hard radiated photons
at the time of the analysis.}

The hadronic channel was investigated in an earlier publication~\cite{OPAL_ggjj}
using LEP1 data from the years 1991 -- 1994.
A sample of 138~\ipb\ events was used in this analysis,
accumulated at energies between 88.28 and 94.28 GeV, and
corresponding to 3.51 million hadronic \Zzero\ decays.
This hadronic channel analysis observed 3 candidates having di-photon
mass greater than 40~GeV with an expected background of $5.4\pm3.0$
events. 
The hadronic channel for an additional 34~\ipb\ of data
was investigated by applying the analysis described
in Section 4.2 to LEP1 data from the year 1995,
corresponding to 0.72 million hadronic \Zzero\ decays.
To reduce the backgrounds at the LEP1 energy, and
to allow for better description of the data by the simulation
programs, the cut on photon energies was modified from the one
used in Section 4.2:
\begin{itemize}
\item ${\rm E}_{\gamma 1} > 15$ GeV, ${\rm E}_{\gamma 2} > 15$ GeV.
\end{itemize}

The hadronic channel 1995 analysis observed one candidate having di-photon
mass greater than 40 GeV, at \mgg\ = 77.2~GeV.
Table~\ref{T:lep1} shows the events passing the cuts of Section 4.2, as well
as the predictions from the simulation programs JETSET 7.4 and HERWIG 5.8.
The efficiencies for Higgs boson signals of several masses are also shown
in the table. 

\section{Results}

\begin{figure}[!p]
        \vspace{0.8cm}
        \begin{center}
            \resizebox{\linewidth}{!}{\includegraphics{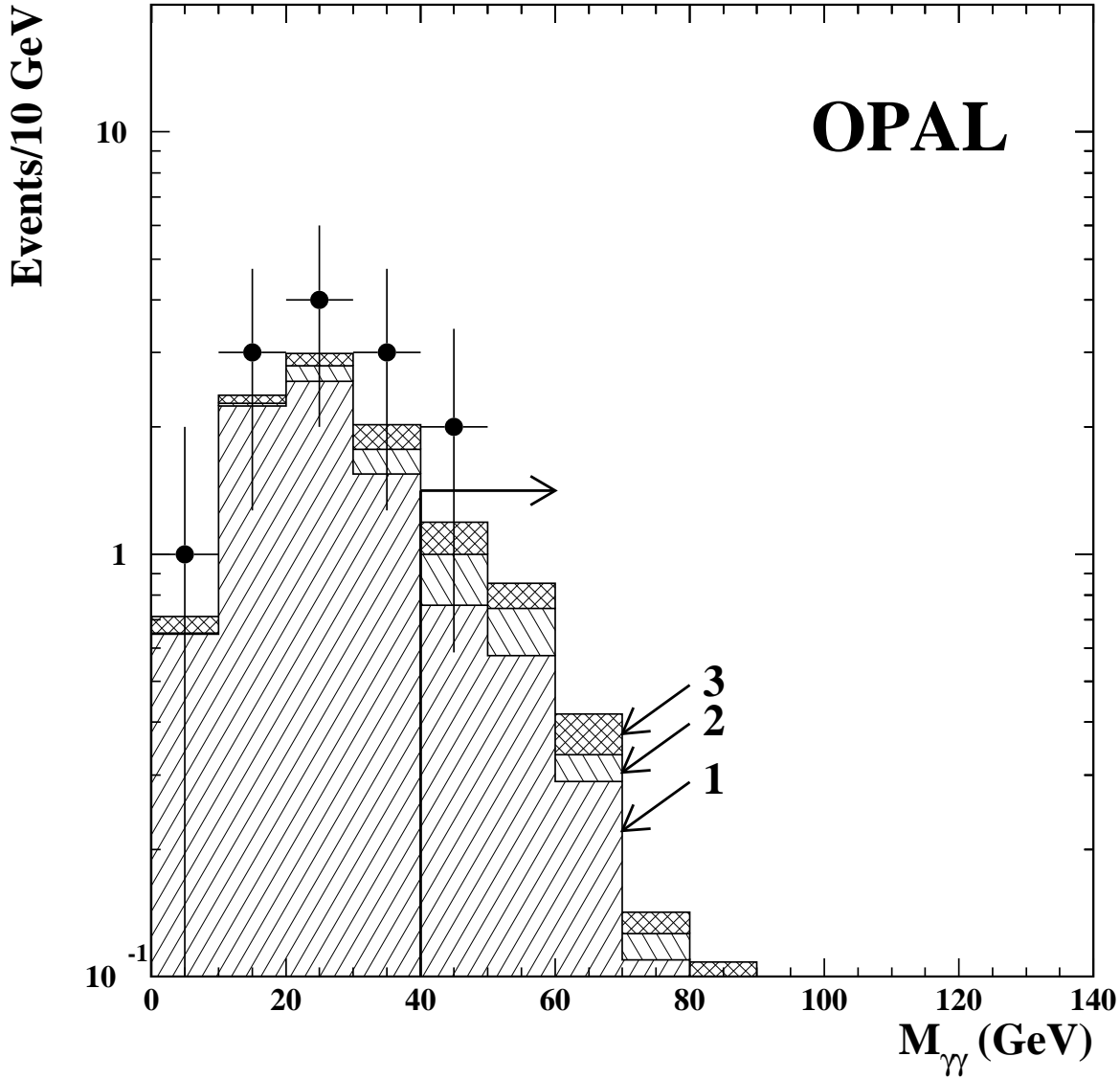} }
        \caption[COMGG]{    
                  Distribution of mass of the two highest energy photons
                  for LEP1.5 and LEP2 events, after all cuts except
                  the one on \mgg; all search channels are included.
                  Data are shown as points with error bars.
                  Background simulation is shown as a histogram with
                  (1) indicating the hadronic search channel,
                  (2) indicating the charged lepton search channel, and
                  (3) indicating the missing energy search channel.
                  The \mgg\ cut is indicated.
        \label{COMGG} }
        \end{center}
    \end{figure}

For the higher energy LEP1.5 and LEP2 data,
the di-photon invariant mass distribution for the events passing all cuts
is shown in Figure~\ref{COMGG};
the simulation of Standard Model backgrounds is also shown in the figure. 
Summing over all expected background sources yields 
$11.7~\pm~0.5$ events expected versus 13 observed. 
The kinematic properties of the candidate events are summarized in
Table \ref{T:parm}. 
Moreover, the qualitative agreement between 
the data and simulation of Standard Model processes is good; therefore 
no new physics process is suggested. 
After requiring a minimum di-photon mass of 40 GeV, 
2 candidates from the LEP1.5 and LEP2 data were left, 
with the missing-energy and hadronic channels each contributing 1 event;
this compares well with the $3.0\pm0.2$ expected from Standard   
Model backgrounds.

The uncertainties pertinent to the limits on production rates
and di-photon branching ratios arose from statistics of the data, 
systematic uncertainty on the luminosity, 
statistical errors on background simulations,
and a systematic error derived
from the level of concordance between backgrounds and their simulation.
The systematic error on the integrated luminosity of the data
(0.6\% for LEP2 energies)
contributed negligibly to the limits.
Statistical uncertainty on the predicted Standard Model background
was dominated by the PYTHIA sample, for which 3000 \ipb\
was generated at the LEP2 energies.
After the cuts on $\theta_{\gamma}$ and $E_{\gamma}$, which effectively
removed the 4-fermion and 2-photon backgrounds, the remaining background
was modelled very well by PYTHIA, as demonstrated in Figure~\ref{X1}.
%
The systematic uncertainty on the background modelling was assessed
by varying the cuts by one standard deviation on the 
experimental resolution of the quantity involved.
The cuts on photon energies are very robust; uncertainties in
electromagnetic cluster energies contribute negligibly to the systematic
error.  The cut most sensitive to background simulation and detector
resolution is that on the photon polar angles.
The method of cuts variation gives a possible increase in expected
backgrounds smaller that the statistical error on the simulation
datasets, approximately 0.2 events in the hadronic channel for the
LEP2 data.
%
The same cut-variation technique applied to the efficiency for
an expected signal yields a contribution to the systematic
uncertainty which is much smaller than the uncertainty from
simulation statistics.

\begin{figure}[!htb]
        \vspace{0.8cm}
        \begin{center}
            \resizebox{\linewidth}{!}{\includegraphics{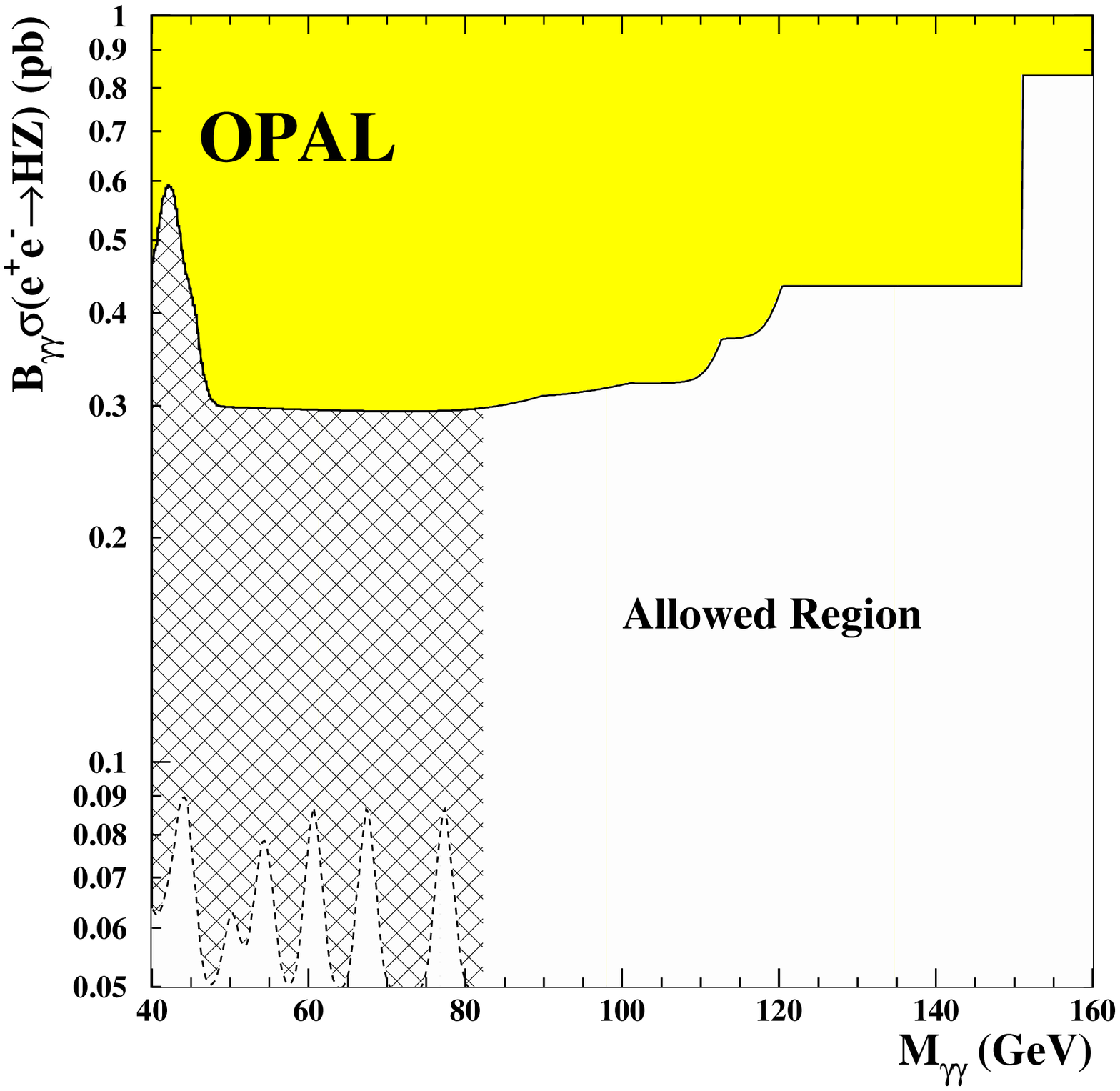} }
        \caption[bslim]{
                 95\% Confidence Level Upper Limit on 
                 $B$($\Hboson \ra \gaga)\times\sigma(\epem\ra\Hboson\Zboson)$.
                Solid curve represents the LEP2 limit for 
                $\roots = 161-172$ GeV; 
                the shaded region is excluded.
                The cross-hatched region is excluded by the LEP1 analysis for
                $\roots \approx 91$ GeV.
        \label{bslim} }
        \end{center}
    \end{figure}

From the events passing the cuts, the 95\% C.L. upper limit (CLUL) 
on the number of signal events
at a given di-photon mass was computed using the method of Bock~\cite{BOCK}.
The method
introduced for every candidate event a weight based on the di-photon mass
resolution and the branching fraction of the \Zzero\ final state. 
A mass-dependent 95\% confidence level upper limit
based on the total weight-sum of all candidate events was computed.
The expected backgrounds were not subtracted in computing
the 95\% CLUL; this results in conservative upper limits.
Furthermore, when the statistical method 
of Bock is used to present the results,
where each candidate event weakens the CLUL only in
the vicinity of its mass,
very little degradation in the upper limits is seen.
The results, in the form of upper limits on production
cross section (times di-photon branching fraction) 
are shown in Figure~\ref{bslim};
because the energies are similar,
the 161~GeV and 172~GeV data have been combined.
In computing these limits, the efficiency was set to 0
for recoil masses less than 10 GeV 
because of uncertainty in simulating the fragmentation process
at such low jet energies.
The step-like nature of the limit between di-photon masses of
90 and 120 GeV is due to the
recoil mass cut in the charged lepton channel
and the cut on photon energies in the missing energy channel.
The step at 151 GeV is due to the increase
in kinematic region afforded by the highest energy (172 GeV) data.
The limits
from LEP1 are compared to those obtained at LEP2 energies in
the figure.  
The larger LEP1 event sample affords a better
limit in the di-photon mass range below 85 GeV. The LEP2
events allow for limits up to nearly twice the LEP1 energy.

\begin{figure}[!htb]
        \vspace{0.8cm}
        \begin{center}
            \resizebox{\linewidth}{!}{\includegraphics{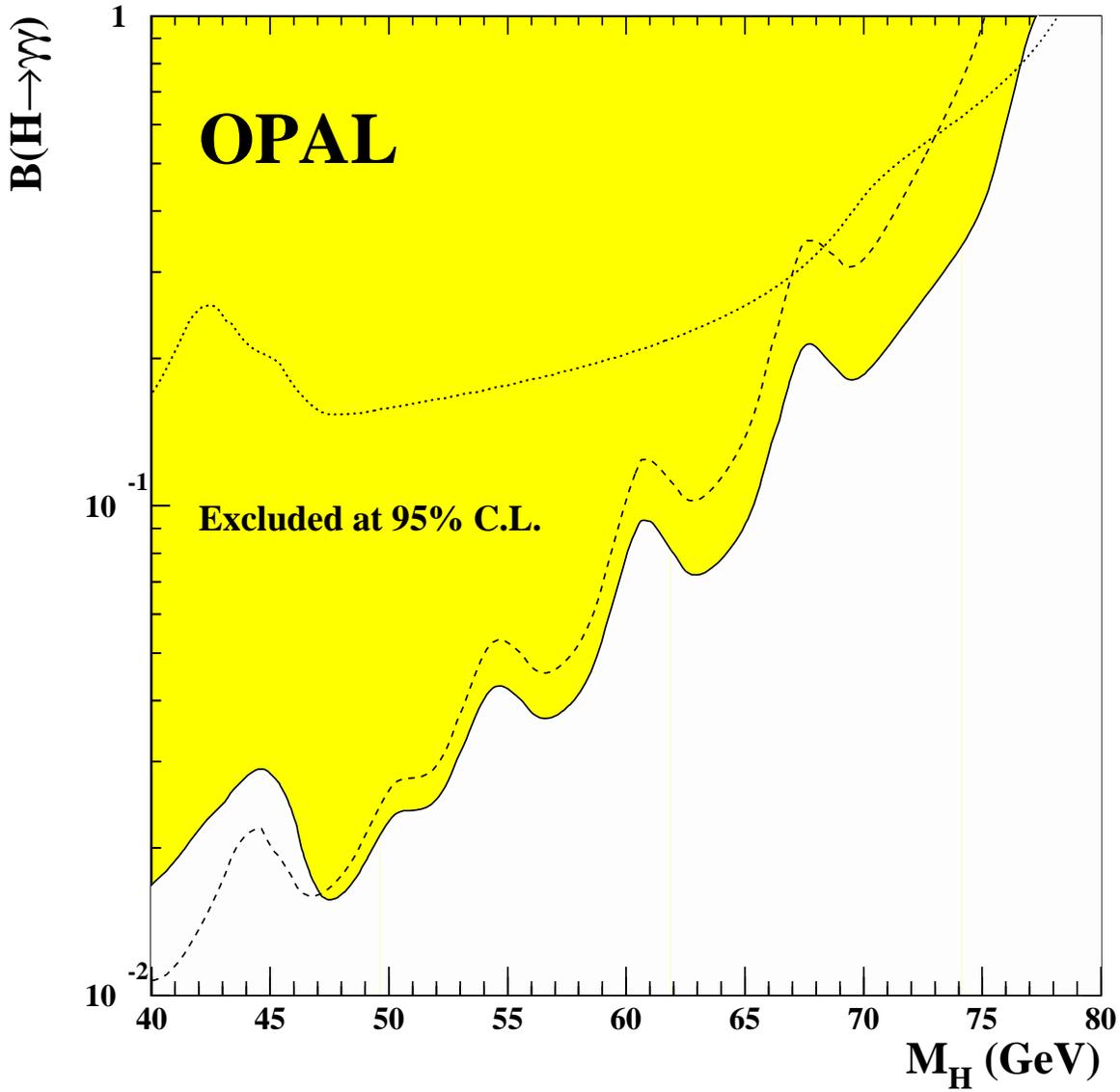} }
        \caption[bgglim]{
                 95\% Confidence Level Upper Limit on $B$($\Hboson \ra \gaga$)
                 for Standard Model Higgs boson production
                 using data from $\roots$ = 91 GeV (dashed line),
                 133, 161 and 172 GeV (dotted line),
                 and all data combined (solid line).
                The shaded region is excluded.
        \label{bgglim} }
        \end{center}
    \end{figure}

To incorporate the $\roots$ dependence among the several centre-of-mass energies,
the Standard Model $\Hboson\Zboson$ production cross section can be factored out
of the limits given in Figure~\ref{bslim}
to set upper limits on the branching fraction for $\Hboson \ra \gaga$
within the context of this model.
This factorization affords a more meaningful presentation of the LEP2
data because of the large phase space factors at $\roots = 161-172$ GeV
(the LEP1.5 data contribute only modestly to these limits
because of the lower energy and small integrated luminosity).
The resulting limits on $B$($\Hboson \ra \gaga$)
are shown in Figure~\ref{bgglim}, where the limits obtained
separately from LEP1, and LEP1.5 and LEP2 combined, are compared.
The LEP1 search had one high mass event at $\mdip = 77.2$ GeV;
this event accounts for the reason the LEP1 data give no useful limit beyond 75 GeV. 
Figure~\ref{bgglim} sets limits on the di-photon branching
fraction up to $\MH = 77$ GeV.

The limit on the Standard Model branching ratio shown in Figure~\ref{bgglim}
can be used to rule out Higgs bosons in certain nonstandard models
in which, unlike the minimal Standard
Model particle, the Higgs boson couples only to bosons.  
In the ``Bosonic" Higgs model~\cite{typeI},
the coupling of the nonstandard Higgs to the \Zzero\
maintains the Standard Model production rate, while the di-photon
branching fraction is larger than 70\% for $\MH < 80$ GeV.
(In some other models~\cite{Akeroyd} the coupling to the \Zzero\
is even larger than the minimal Standard Model value.)
Using the
LEP1, LEP1.5, and LEP2 data, a lower limit of 76.5 GeV is obtained at the
95\% confidence level.

\begin{figure}[!htb]
        \vspace{0.8cm}
        \begin{center}
            \resizebox{\linewidth}{!}{\includegraphics{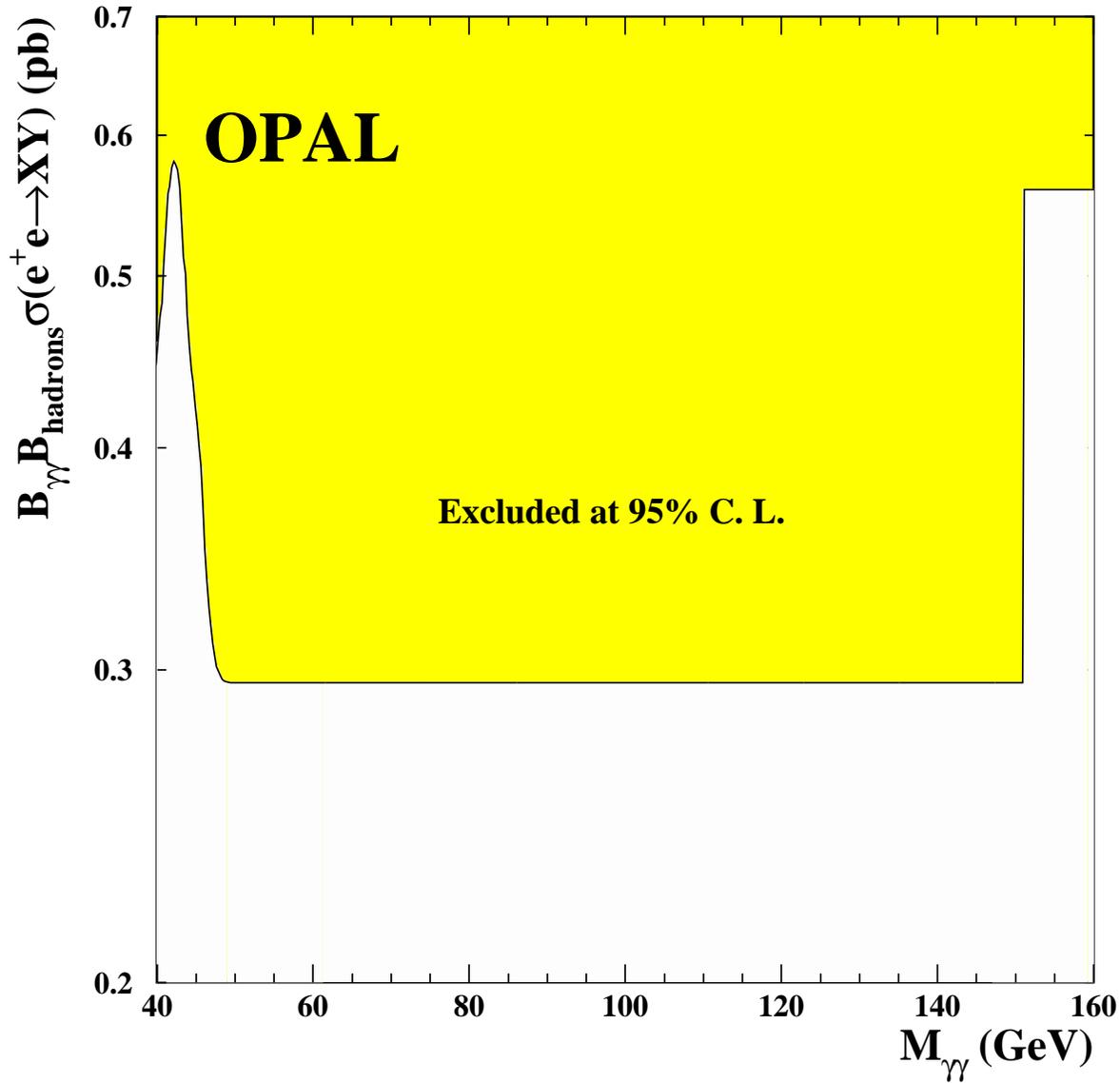} }
        \caption[limxy]{
                 95\% Confidence Level Upper Limit on $B$($\rm X \ra \gaga)\times$
                              $B$(${\rm Y} \ra hadrons)\times$
                              $\sigma(\epem\ra \rm XY)$,
                 for scalar X and vector Y, using the hadronic channel
                 analysis with data from LEP2.    
                The shaded region is excluded.
        \label{limxy} }
        \end{center}
    \end{figure}

More general limits on $\epem \ra \rm XY$ production
can be obtained using the LEP2 hadronic channel alone.
To compute \mgg\ dependent limits, the
PYTHIA and HZHA generators have been used to generate a grid of X
and Y (recoil particle) masses. It was assumed that X was a scalar,
and the cases where Y was a vector or scalar were investigated
(the efficiencies were found to be almost equal for Y scalar or vector). 
Limits  
were computed using the efficiency at a given \mgg\ which was the minimum
for the kinematically allowed variation of $\rm \MY$. 
The limits thus obtained are shown in Figure~\ref{limxy}.

\section{Conclusions}

Using a data sample of 
25.7~\ipb\ taken at centre-of-mass energies from 130 to 172 GeV
and 173~\ipb\ taken near 91 GeV,
a search for a massive di-photon resonance has been performed. 
For \mgg$>40$ GeV, a total of 2 candidates survived
all selection requirements on the LEP1.5 and LEP2 data. 
The number of observed candidates
was consistent with the Standard Model prediction
of $3.0\pm0.2$ background events.
From the LEP2 data,
upper limits on $B$($\Hboson \ra \gaga)\times\sigma(\epem\ra\Hboson\Zboson)$
of 290 -- 830~fb are obtained over $40 < \MH < 160$ GeV.
From the LEP2 hadronic channel alone, an upper limit on
$B$(${\rm X} \ra \gaga)\times B$(${\rm Y} \ra hadrons)\times$$\sigma(\epem\ra {\rm XY})$,
for X a scalar particle,
can be placed at 290 fb over the mass range $50 < \mdip  < 150$ GeV. 
At $\roots \approx 91$~GeV, the LEP1 data upper limit on
$B$($\Hboson \ra \gaga)\times\sigma(\epem\ra\Hboson\Zboson)$
is better than 90~fb for $40 < \mdip  < 80$ GeV.
Data from $\roots \approx 91$ GeV can be combined with the LEP1.5 and LEP2 data;
these combined data can be interpreted within the context of the Standard Model
to set a limit on $B$($\Hboson \ra \gaga$) up to a Higgs boson mass of
77 GeV, provided the Higgs particle is produced via $\epem \ra \Hboson \Zboson$.
A lower mass bound of 76.5 GeV is set at the 95\% confidence level for
Higgs particles which couple only to gauge bosons but still couple to
the \Zzero\ at minimal Standard Model strength.

\bigskip
\noindent {\Large \bf Acknowledgements}
\par
\noindent We particularly wish to thank the SL Division for the efficient operation
of the LEP accelerator at all energies
 and for
their continuing close cooperation with
our experimental group.  We thank our colleagues from CEA, DAPNIA/SPP,
CE-Saclay for their efforts over the years on the time-of-flight and trigger
systems which we continue to use.  In addition to the support staff at our own
institutions we are pleased to acknowledge the  \\
Department of Energy, USA, \\
National Science Foundation, USA, \\
Particle Physics and Astronomy Research Council, UK, \\
Natural Sciences and Engineering Research Council, Canada, \\
Israel Science Foundation, administered by the Israel
Academy of Science and Humanities, \\
Minerva Gesellschaft, \\
Benoziyo Center for High Energy Physics,\\
Japanese Ministry of Education, Science and Culture (the
Monbusho) and a grant under the Monbusho International
Science Research Program,\\
German Israeli Bi-national Science Foundation (GIF), \\
Bundesministerium f\"ur Bildung, Wissenschaft,
Forschung und Technologie, Germany, \\
National Research Council of Canada, \\
Hungarian Foundation for Scientific Research, OTKA T-016660,
T023793 and OTKA F-023259.\\

\newpage

\newpage


\begin{table}[!htbp]
  \begin{center}
    \begin{tabular}{|l||r||r|r|r|r||r|r|}\hline

    Cut  &Data & $\Sigma$Bkgd &$(\gamma/{\rm Z})^{\ast}$ & \multicolumn{1}{|c|}{4f} 
                                & $\epem\qqbar$ &$\MH=40$ &$\MH=70$  \\ \hline
\hline
\multicolumn{8}{|c|}{133 GeV} \\ \hline
    Multiplicity                &1553 &1557. &1529. &16.1  &12.0 &0.99 &0.97 \\ \hline
    Precuts                     &736  &804.  &794.  &9.93  &0.00 &0.91 &0.88 \\ \hline
    $N_{\gamma}\ge2$            &16   &10.4  &10.3  &0.13  &0.00 &0.60 &0.65 \\ \hline
    $\cos\theta_{\gamma}$ cut   &6    &3.75  &3.75  &0.00  &0.00 &0.44 &0.51 \\ \hline
\hline
\multicolumn{8}{|c|}{161 GeV} \\ \hline
    Multiplicity                &1525 &1432. &1346. &55.2 &30.6   &0.99 &0.99 \\ \hline
    Precuts                     &523  &511.  &480.  &30.7 &0.53   &0.87 &0.93 \\ \hline
    $N_{\gamma}\ge2$            &10   &7.95  &7.83  &0.12 &0.00   &0.61 &0.62 \\ \hline
    $\cos\theta_{\gamma}$ cut   &3    &2.51  &2.51  &0.00 &0.00   &0.51 &0.49 \\ \hline
\hline
\multicolumn{8}{|c|}{172 GeV} \\ \hline
    Multiplicity                &1409 &1280. &1126.  &126.  &28.9   &0.99 &0.99  \\ \hline
    Precuts                     &461  &465.  &386.   &78.5  &0.30   &0.87 &0.92  \\ \hline
    $N_{\gamma}\ge2$            &7    &6.65  &6.64   &0.01  &0.00   &0.67 &0.60  \\ \hline
    $\cos\theta_{\gamma}$ cut   &1    &2.00  &2.00   &0.00  &0.00   &0.53 &0.49  \\ \hline
    \hline\end{tabular}\newline
  \end{center}
  \caption[Data and MC after cuts]
  {Events remaining in the LEP1.5 and LEP2 hadronic channel search after cumulative cuts indicated.
   The background simulation samples are 
   scaled to 
   5.4 \ipb\ for $\sqrt{s}$ = 133 GeV,
   10.0 \ipb\ for $\sqrt{s}$ = 161 GeV, and
   to 10.3 \ipb\ for $\sqrt{s}$ = 172 GeV.
   In addition to the total simulated background, the simulations for
   $(\gamma/{\rm Z})^{\ast}$, 4-fermion (``4f"), and Two-photon ($\epem\qqbar$) states
   are shown.
   ``$\MH=40$" and ``$\MH=70$" indicate the efficiency for simulated \Hzero\Zzero\ events
   with the Higgs mass equal to 40 and 70 GeV, respectively.}
  \label{T:qq1}

\end{table}


\begin{table}[!htbp]                      
\begin{center}
\begin{tabular}{|l||r||r|r|r|r|r||r|r|}\hline
 Cut                 & Data   &$\Sigma$Bkgd & $\epem$  & $ \tptm$  & $\mpmm$ & \eeff\  
                                            & $\MH=40$ & $\MH=70$  \\ \hline 
\multicolumn{9}{|c|}{135 GeV} \\ \hline
$\ell\ell\gaga$ presel. & 395  & 179.  &  55.4  &   49.6  &  5.94  & 68.3  & 0.78 & 0.80  \\ \hline
$N_{\gamma}\geq 2$      &   2  & 3.96  &  2.45  &   0.92  &  0.56  & 0.03  & 0.59 & 0.61  \\ \hline
$N_{\mrm jet}\geq 2$    &   2  & 2.60  &  1.36  &   0.72  &  0.52  & 0.01  & 0.58 & 0.58  \\ \hline
 \Lgg\                  &   0  & 1.41  &  0.65  &   0.43  &  0.33  & 0.01  & 0.47 & 0.48  \\ \hline
$\Mrec$                 &   0  & 0.68  &  0.13  &   0.29  &  0.25  & 0.01  & 0.35 & 0.04  \\ \hline
\multicolumn{9}{|c|}{161 GeV} \\ \hline
$\ell\ell\gaga$ presel. & 434  & 183.  &  72.8  &  46.8  &   6.73  & 57.1  & 0.77  & 0.81  \\ \hline
$N_{\gamma}\geq 2$      &   5  &5.28   &  3.44  &  0.94  &   0.80  & 0.10  & 0.62  & 0.67  \\ \hline
$N_{\mrm jet}\geq 2$    &   1  &3.21   &  1.70  &  0.67  &   0.74  & 0.10  & 0.60  & 0.65  \\ \hline
\Lgg                    &   0  &1.33   &  0.61  &  0.33  &   0.39  & 0.00  & 0.49  & 0.53  \\ \hline
$\Mrec$                 &   0  &0.60   &  0.24  &  0.15  &   0.21  & 0.00  & 0.46  & 0.48  \\ \hline
\hline
\multicolumn{9}{|c|}{172 GeV} \\ \hline
$\ell\ell\gaga$ presel. & 323  &173.   &  67.8  &  39.9  &   5.80  & 59.1  & 0.77 & 0.80 \\ \hline
$N_{\gamma}\geq 2$      &   5  &4.70   &  2.74  &  0.83  &   0.69  & 0.43  & 0.61 & 0.60 \\ \hline
$N_{\mrm jet}\geq 2$    &   1  &2.65   &  1.08  &  0.61  &   0.63  & 0.33  & 0.58 & 0.58 \\ \hline
\Lgg\                   &   0  &1.17   &  0.42  &  0.32  &   0.32  & 0.10  & 0.49 & 0.47 \\ \hline
$\Mrec$                 &   0  &0.36   &  0.08  &  0.14  &   0.15  & 0.00  & 0.43 & 0.45 \\ \hline
\hline
\end{tabular}
\end{center}
  \caption[Data and MC after cuts]
  {Events remaining after cumulative cuts indicated, for the 
   LEP1.5 and LEP2 leptonic channel analysis. 
   The row denoted ``$\ell\ell\gaga$ presel." refers
   to the combined general low-multiplicity selection and the precuts described
   in Section \ref{s:llgg}. 
   The contributions from $\epem$-pair,
   $\mu$-pair, $\tau$-pair production and \eeff\ final states
   determined from background simulations are shown. The simulated datasets 
   have been normalized to 
   5.4 \ipb\ for $\sqrt{s}$ = 133 GeV,
   10.0 \ipb\ for $\sqrt{s}$ = 161 GeV, and
   to 10.3 \ipb\ for $\sqrt{s}$ = 172 GeV. 
   Also shown is the acceptance for a Higgs signal for 
   40 and 70 GeV mass denoted as columns ``$\MH=40$" and ``$\MH=70$", respectively.
The poor agreement between data and background simulations in the preselection
category results from inadequate modelling of material near the beampipe in the
forward region.
}  \label{T:ggll}
\end{table}


\begin{table}[!htbp]
\begin{center}
\begin{tabular}{|l||r||r||r|r|r|r|r||r|r|}\hline
 Cut  &   Data  &$\Sigma$Bkgd&  $\epem$  & $\nunu\gaga$   & $\gaga$ & $\ell^+\ell^-$ & \eeff\  
                                                    & $\MH=40$ & $\MH=70$ \\ \hline
\multicolumn{10}{|c|}{133 GeV} \\ \hline
$\nn\gaga$ presel.
         & 32 & 1.73  &  0.19  &     0.74  &     0.80  &     0.01  &  0.00  & 0.68  & 0.64  \\ \hline
 \TPONE\ & 4  & 1.26  &  0.06  &     0.74  &     0.45  &     0.01  &  0.00  & 0.67  & 0.63  \\ \hline
 \TPTWO\ & 2  & 0.90  &  0.06  &     0.74  &     0.10  &     0.01  &  0.00  & 0.67  & 0.63  \\ \hline
 \Lgg\   & 0  & 0.59  &  0.00  &     0.54  &     0.05  &     0.00  &  0.00  & 0.58  & 0.55  \\ \hline
\hline
\multicolumn{10}{|c|}{161 GeV} \\ \hline
 $\nn\gaga$ presel.
         & 32 & 8.02 &  5.49  &     0.98  &     1.30  &     0.04  &  0.21  & 0.65 & 0.70 \\ \hline
 \TPONE\ &  5 & 4.85 &  3.21  &     0.98  &     0.64  &     0.02  &  0.00  & 0.64 & 0.70 \\ \hline
 \TPTWO\ &  1 & 2.01 &  0.73  &     0.98  &     0.29  &     0.01  &  0.00  & 0.63 & 0.69 \\ \hline
 \Lgg\   &  1 & 0.63 &  0.00  &     0.55  &     0.08  &     0.00  &  0.00  & 0.54 & 0.58 \\ \hline
\hline
\multicolumn{10}{|c|}{172 GeV} \\ \hline
 $\nn\gaga$ presel.
         & 27 & 7.88 &  4.83  &     0.93  &     1.15  &     0.03  &  0.93  &  0.62 & 0.72  \\ \hline
 \TPONE\ &  5 & 4.54 &  2.80  &     0.93  &     0.59  &     0.01  &  0.21  &  0.61 & 0.72  \\ \hline
 \TPTWO\ &  3 & 1.48 &  0.32  &     0.93  &     0.23  &     0.01  &  0.00  &  0.60 & 0.71  \\ \hline
 \Lgg\   &  2 & 0.54 &  0.00  &     0.44  &     0.10  &     0.00  &  0.00  &  0.53 & 0.62 \\ \hline
\hline
\end{tabular}
\end{center}

  \caption[Data and MC after cuts]
  {Events remaining after cumulative cuts indicated for the 
   LEP1.5 and LEP2 missing energy channel search. 
   The row denoted ``$\nn\gaga$ presel." refers
   to the combined general low-multiplicity selection and the precuts described
   in Section \ref{s:nngg}. 
   The contributions from $\epem$-pair,
   $\nunu\gaga$, $\gaga$, lepton pair ($\ell\equiv\mu,\tau$) production and \eeff\ final states
   determined from background simulations are shown. The simulation datasets
   have been normalized to 
   5.4 \ipb\ for $\sqrt{s}$ = 133 GeV,
   10.0 \ipb\ for $\sqrt{s}$ = 161 GeV, and
   to 10.3 \ipb\ for $\sqrt{s}$ = 172 GeV.
   Also shown is the acceptance for a Higgs signal for 
   40 and 70 GeV mass denoted as columns ``$\MH=40$" and ``$\MH=70$", respectively.
The poor agreement between data and background simulations in the preselection
category results from inadequate modelling of material near the beampipe in
the forward region.
}  \label{T:ggnn}
\end{table}

\begin{table}[!htbp]
  \begin{center}
    \begin{tabular}{|l||r|r|r||r|r|}\hline

    Cut  &Data &JETSET 7.4  &HERWIG 5.8    &$\MH=40$ &$\MH=70$  \\ \hline \hline
    Multiplicity   &720432      &720432  &720432   &0.99 &0.91 \\ \hline
    Precuts                     &469235  &471123.  &465251          &0.94 &0.85 \\ \hline
    $N_{\gamma}\ge2$            &13      &5.13     &4.00            &0.50 &0.61 \\ \hline
    $\cos\theta_{\gamma}$ cut   &2       &3.67     &2.00            &0.38 &0.45 \\ \hline
    \hline\end{tabular}\newline
  \end{center}
  \caption[Data and MC after cuts]
  {Events remaining in the 1995 LEP1 hadronic channel search after cumulative cuts indicated.
   The background predictions for the JETSET 7.4 and HERWIG 5.8 
   $(\gamma/{\rm Z})^{\ast}$ simulations are shown; the simulations have been scaled
   to the number of multihadrons in the data passing the multiplicity cut.
   ``$\MH=40$" and ``$\MH=70$" indicate the efficiency for simulation of \Hzero\Zzero\ events
   with the Higgs mass equal to 40 and 70 GeV, respectively.
   The poor agreement between data and simulations at the $N_{\gamma}\ge2$ cut
   is due to poor modelling of isolated $\pi^0$ mesons.}
  \label{T:lep1}

\end{table}

\begin{table}[!htbp] 
 \begin{center}
 \begin{tabular}{|c|c|c|c|c|c|c|c|}\hline
Channel & $\roots$(GeV) & \mgg(GeV) & $M_{\rm recoil}$ (GeV)  
 & ${\rm E}_{\gamma1}$ (GeV)& $\cos\theta_{\gamma1}$& ${\rm E}_{\gamma2}$ (GeV)&$\cos\theta_{\gamma2}$ \\ \hline

$\nn$       &172          &$44.9\pm1.2$ & 93.1  &48.7  &$0.90 $ &18.0  &$-0.27$  \\ \hline
$\qqbar$    &161          &$42.2\pm1.8$ & 79.9  &39.2  &$-0.04$ &27.0  &$-0.81$  \\ \hline
$\nn$       &172          &$39.9\pm3.0$ & 92.6  &51.1  &$0.05 $ &14.7  &$0.79 $ \\ \hline
$\qqbar$    &172          &$36.8\pm1.4$ & 90.4  &60.3  &$0.63 $ &5.8   &$-0.31$  \\ \hline
$\qqbar$    &130          &$31.0\pm1.5$ & 89.4  &23.7  &$-0.35$ &14.4  &$10.71$  \\ \hline
$\qqbar$    &130          &$28.7\pm1.0$ & 91.7  &24.9  &$0.51 $ &11.2  &$0.51 $ \\ \hline
$\qqbar$    &130          &$25.9\pm1.1$ & 86.0  &31.0  &$0.57 $ &8.3   &$0.61 $ \\ \hline
$\qqbar$    &161          &$24.9\pm1.0$ & 72.1  &54.5  &$0.66 $ &11.8  &$0.63 $ \\ \hline
$\qqbar$    &136          &$22.5\pm1.0$ & 82.0  &40.7  &$0.64 $ &4.6   &$0.42 $ \\ \hline
$\nn$       &161          &$15.8\pm0.5$ & 106.9 &32.4  &$0.58 $ &13.4  &$0.10 $ \\ \hline
$\qqbar$    &136          &$13.5\pm1.1$ & 109.7 &19.6  &$-0.74$ &5.1   &$0.36 $ \\ \hline
$\qqbar$    &161          &$12.1\pm0.5$ & 85.6  &53.1  &$0.60 $ &5.1   &$-0.10$  \\ \hline
$\qqbar$    &130          &$6.4\pm0.2$  & 63.4  &34.1  &$-0.47$ &15.8  &$-0.21$  \\ \hline
\hline\end{tabular}\newline
\end{center}
\caption[Masses and Energies of Candidate Events.]
{Masses and energies of candidate events from the LEP1.5 and LEP2 searches,
after all cuts except the one on di-photon mass.
The events are ordered by di-photon mass.}
\label{T:parm}
\end{table}



\newpage

\newpage

\newpage


\newpage

\newpage

\newpage

\newpage

\newpage

\newpage

\newpage

\end{document}